\documentclass[11pt]{article}
\newif\ifnotconf
\notconffalse
\def\showauthornotes{0}
\def\showkeys{0}
\def\showdraftbox{0}
\usepackage{amsmath,amsfonts,amssymb,amsthm,mathtools,enumerate}
\usepackage{paralist}
\usepackage{enumitem}
\usepackage{graphicx}
\usepackage{lmodern}
\usepackage{tikz}
\usetikzlibrary{arrows,calc,positioning}
\usepackage{nicefrac}
\usepackage{listings}
\usepackage{xcolor,xspace}
\usepackage[most]{tcolorbox}
\usepackage[T1]{fontenc}
\usepackage{dsfont}
\usepackage{bbm}
\usepackage{newpxmath} %%Palladio clone for math fonts

\usepackage{newpxtext} %%Palladio clone -- new implementation
\usepackage{url}%this line and the next are related to hyperlinks
\usepackage{hyperref}
\hypersetup{
    colorlinks=true,   % Make the links colored
    pdfstartview=Fit, 
    linkcolor=blue,    % Color for section, figure, table, etc. links
    citecolor=magenta, % Color for bibliography citations
    urlcolor=green,     % Color for URLs (e.g., in bibliography)
}
\usepackage[
  hyperref=true,      	% Link to online documents
  backend=biber,      % Bibliography backend
  sorting=nyt,       	% Sorts by (name, year, title)
  style=alphabetic, 	% Citations look like [Har77]
  maxalphanames=4, % Maximum authors to list out fully in the label before using "+"
  minalphanames=3,  % How many author initials to keep if it does truncate
  maxbibnames=999, % How many authors full name can appear in Bibliography
  minbibnames=999
]{biblatex}

\usepackage[capitalise,nameinlink]{cleveref}

\crefname{ineq}{Ineq.}{Inequalities}
\creflabelformat{ineq}{#2{\upshape(#1)}#3}
\crefname{alg}{Algorithm}{Algorithms}
\creflabelformat{alg}{#2{\upshape(#1)}#3}
\crefname{fact}{Fact}{Facts}
\creflabelformat{fact}{#2{\upshape(#1)}#3}
\makeatletter
\AddToHook{cmd/appendix/before}{\crefalias{section}{appendix}}
\makeatother

\usepackage{thmtools}
\usepackage{thm-restate}
\usepackage{float}
\floatstyle{boxed}
\restylefloat{figure}
\usepackage{framed}

\ifnum\showkeys=1
\usepackage[color]{showkeys}
\fi

\DeclareMathOperator{\Span}{\operatorname{span}}

\renewcommand{\bar}[1]{\ensuremath{\overline{#1}}}

\newcommand{\1}[1]{\mathds{1}_{\left[#1\right]}}

\ifnum\showauthornotes=1
\newcommand{\Authornote}[2]{{\sf\small\color{red}{[#1: #2]}}}
\newcommand{\Authorcomment}[2]{{\sf \small\color{gray}{[#1: #2]}}}
\newcommand{\Authorfnote}[2]{\footnote{\color{red}{#1: #2}}}
\else
\newcommand{\Authornote}[2]{}
\newcommand{\Authorcomment}[2]{}
\newcommand{\Authorfnote}[2]{}
\fi
\newtheorem{theorem}{Theorem}[section]

\newtheorem{lemma}[theorem]{Lemma}

\newtheorem{corollary}[theorem]{Corollary}

\usepackage{algorithm}
\usepackage{algorithmic}
\def\FullBox{\hbox{\vrule width 6pt height 6pt depth 0pt}}

\def\qed{\ifmmode\qquad\FullBox\else{\unskip\nobreak\hfil
\penalty50\hskip1em\null\nobreak\hfil\FullBox\parfillskip=0pt\finalhyphendemerits=0\endgraf}\fi}

\def\qedsketch{\ifmmode\Box\else{\unskip\nobreak\hfil
\penalty50\hskip1em\null\nobreak\hfil$\Box$
\parfillskip=0pt\finalhyphendemerits=0\endgraf}\fi}

\ifnotconf%
  \renewenvironment{proof}{\bgroup\trivlist\item {\bf Proof:\enspace}}{\qed\endtrivlist\egroup}
\fi

\newcommand{\mper}{\,.}
\newcommand{\mcom}{\,,}

\newcommand{\vect}[1]{\mathbf{#1}}
\newcommand{\innerprod}[2]{\ensuremath{\left\langle #1, #2 \right\rangle}}

\newcommand{\Sch}{\mathfunction{\mathsf{S}~\!}}
\makeatletter
\def\norm#1{%fancy norm command
  \@ifnextchar\bgroup%
   {\normalpnorm{#1}}%
   {\defaultnorm{#1}}%
}
\def\defaultnorm#1{%
    \renderNorm{#1}{}
}
\def\normalpnorm#1#2{%
   \@ifnextchar\bgroup%
   {\banaxnorm{#1}{#2}}%
   {\renderNorm{#2}{#1}}%\ensuremath{\left\lVert {#2} \right\rVert_{{#1}}}}%
}

\def\banaxnorm#1#2#3{%
    \renderNorm{#3}{#1\to#2 }
}

\def\renderNorm#1#2{%
    \@ifnextchar^%
    {\fixExponent{#1}{#2}}%
    {\ensuremath{\mathchoice%
        {\lVert #1 \rVert_{#2}}%
        {\lVert #1 \rVert_{#2}}%
        {\lVert #1 \rVert_{#2}}%
        {\lVert #1 \rVert_{#2}}}%
    }%
}
\def\fixExponent#1#2^#3{%
    \ensuremath{{\lVert #1 \rVert^{#3}_{#2}}}%
}

\def\enorm#1#2{%
   \@ifnextchar\bgroup%
   {\norm{L_{#1}}{L_{#2}}}%
   {\norm{L_{#1}}{#2}}%
}

\def\cnorm#1#2{%
   \@ifnextchar\bgroup%
   {\norm{\ell_{#1}}{\ell_{#2}}}%
   {\norm{\ell_{#1}}{#2}}%
}

\def\schnorm#1#2{%
   \@ifnextchar\bgroup%
   {\norm{\Sch{#1}}{\Sch{#2}}}%
   {\norm{\Sch{#1}}{#2}}%
}

\makeatother

\makeatletter

\def\HierarchyRender#1{
  \@ifnextchar\bgroup%
  {\hierarchywithlvl{#1}}
  {\nameofhierarchy{#1}}
}
\def\nameofhierarchy#1{#1\xspace}
\def\hierarchywithlvl#1#2{%
    \ensuremath{#1^{(#2)}}\!\xspace
}
\makeatother
\newcommand{\Esymb}{\mathbb{E}}
\newcommand{\Psymb}{\mathbb{P}}

\makeatletter
\def\Pr#1{%
    \ProbabilityRender{\Psymb}{#1}%
}

\def\Ex#1{%
    \ProbabilityRender{\Esymb}{#1}%
}

\def\ProbabilityRender#1#2{%fancy probability command
  \@ifnextchar\bgroup%
  {\renderwithdist{#1}{#2}}
   {\singlervrender{#1}{#2}}
}
\def\singlervrender#1#2{%
   \ensuremath{\mathchoice
       {{#1}\left[ #2 \right]}
       {{#1}[ #2 ]}
       {{#1}[ #2 ]}
       {{#1}[ #2 ]}
   }
}
\def\renderwithdist#1#2#3{%
   \@ifnextchar\bgroup%
   {\superfancyrender{#1}{#2}{#3}}
   {\ensuremath{\mathchoice%
      {\underset{#2}{#1}\left[ #3 \right]}
      {{#1}_{#2}[ #3 ]}
      {{#1}_{#2}[ #3 ]}
      {{#1}_{#2}[ #3 ]}
     }
   }
}
\def\superfancyrender#1#2#3#4#5{
   \ensuremath{\mathchoice
      {\underset{#1}{{#1}}\left#4 #3 \right#5}
      {{#1}_{#2}#4 #3 #5}
      {{#1}_{#2}#4 #3 #5}
      {{#1}_{#2}#4 #3 #5}
   }
}
\makeatother

\makeatletter
\def\mathfunction#1{%fancy math functions: #1 function symbol w/ sub & superscript #2 function argument
  \@ifnextchar\bgroup%
   {\funcWithArg{#1}}%
   {\justFunc{#1}}%
}
\def\justFunc#1{%
  \ensuremath{#1}\xspace
}
\def\funcWithArg#1#2{%
  \@ifnextchar\bgroup%
  {\funcWArgSub{#1}{#2}}%
  {\ensuremath{\mathchoice%
      {#1\!\left(#2\right)}%
      {#1\!\left(#2\right)}%
      {#1(#2)}%
      {#1(#2)}\xspace}%
  }%
}
\def\funcWArgSub#1#2#3{%
  \ensuremath{\mathchoice%
    {#1_{#2}\!\left(#3\right)}%
    {#1_{#2}\!\left(#3\right)}%
    {#1_{#2}(#3)}%
    {#1_{#2}(#3)}
  \xspace}%
}
\makeatother

\DeclareMathOperator{\val}{val}

\DeclareMathOperator{\opt}{opt}
\DeclareMathOperator{\thr}{thr}

\ifnum\showdraftbox=1

\else

\fi

\usetikzlibrary{decorations.markings,arrows,positioning,automata}
\tikzset{->-/.style={decoration={markings,mark=at position .5 with {\arrow{>}}},postaction={decorate}}}

\hypersetup{pdftitle={Inverse-Linear and Inverse-Square Gains over Threshold Rounding at Bounded Degree: Reasonable Boolean 2-CSPs and Beyond},pdfauthor={Mrinalkanti Ghosh}}
\title{\Large Inverse-Linear and Inverse-Square Gains over Threshold Rounding\\
at Bounded Degree: Reasonable Boolean 2-CSPs and Beyond}
\author{Mrinalkanti Ghosh\thanks{%
    Indian Institute of Science, $\quad$
    \texttt{mrinalkantig@iisc.ac.in}, $\quad$
    Funding Source: Walmart Postdoc Fellowship}
  }
\begin{document}
\maketitle
\begin{abstract}
When each variable occurs in at most $d$ constraints, local improvements can
raise the approximation ratio of semidefinite rounding. For weighted Max-Cut
and Max-2-Lin(2), the known gain over the worst-case ratio is
$\Omega(1/(d\log d))$~\cite{BGHKLMM26}. For reasonable SDP solutions of binary
Boolean CSPs, earlier work gives $a+\Omega(1/(d^4(\log d)^{7/2}))$~\cite{GHLM26},
where $a$ is the expected fraction of weight satisfied by Gaussian rounding
with a fixed bounded threshold function of the SDP bias. For weighted Max
2-SAT, the same work gives a gain of
$\Omega(1/(d^2\log d))$ over the worst-case ratio
$\alpha_{\rm LLZ}\simeq0.94016567$.

We give a randomized polynomial-time algorithm that achieves $a+\Omega(1/d^2)$
under a geometric condition on the SDP solution and thresholds that strictly
extends reasonableness. To improve a baseline approximation ratio, it suffices
that the binary constraints near equality satisfy this condition jointly. A separate condition
on these constraints gives an inverse-linear gain. It requires bounded
rounding thresholds, score correlations bounded away from $\pm1$, and oriented
thresholds that have one sign at each variable and stay uniformly away from
zero. For weighted Max 2-SAT this gives the approximation ratio
$\alpha_{\rm LLZ}+\Omega(1/d)$. The same argument gives a gain of $\Omega(1/d)$
over the respective worst-case ratios for Max $\{1,2\}$-Horn-SAT and positive
binary disjunctions with unit clauses of either sign. All three baseline ratios
are sharp under the Unique Games Conjecture~\cite{BHZ23}.
\end{abstract}

\begingroup
\small
\newpage
\tableofcontents
\newpage
\endgroup

\section{Introduction}
\label{sec:intro}

A Boolean 2-CSP has variables taking values in $\{\pm 1\}$. Each constraint
applies a predicate to at most two variables and is satisfied if their labels
form an accepting input. The allowed predicates form the \emph{constraint
language}. In a weighted instance, each constraint has a positive weight. We
seek an assignment that maximizes the total weight of the constraints it
satisfies. Write $\opt$ for this maximum. A Max-Cut edge requires its endpoints
to receive different labels. In Max 2-SAT, the constraints are disjunctions of
at most two literals.

A standard approach solves a semidefinite programming (SDP) relaxation and
applies Gaussian threshold rounding. The SDP solution gives a direction for
each variable. We draw one standard Gaussian vector and take its inner product
with each direction. A variable receives label $+1$ if its score exceeds its
threshold and $-1$ otherwise. For Max-Cut, we may choose an optimal SDP
solution with all variable biases zero. Hyperplane rounding of these vectors
at threshold zero gives a cut of expected weight at least
$\alpha_{\rm GW}\simeq0.87857$ times the optimum~\cite{GW95}. For Max 2-SAT,
choosing the threshold as a function of the variable's mean label in the SDP
solution gives the approximation ratio
$\alpha_{\rm LLZ}\simeq0.94016567$~\cite{LewinLZ02,BHZ23}. Under the Unique
Games Conjecture (UGC), both worst-case ratios are optimal for polynomial-time
algorithms~\cite{KKMO07,BHZ23}.

Can SDP rounding do better on structured instances, particularly when every
variable has bounded degree? We measure \emph{occurrence degree}. A constraint
contributes one to the degree of each variable it depends on, regardless of its
weight. Replacing a constraint of weight $w$ by $k$ copies of weight $w/k$
leaves the objective unchanged but counts as $k$ occurrences at each variable.
If every variable has degree at most $d$, changing one label affects at most
$d$ terms in the objective. This locality gives a gain over the worst-case
ratio for Max-Cut and, more generally, for Max-2-Lin(2), whose constraints ask
for equality or disequality~\cite{FKL02,HK23,BGHKLMM26}. Flipping a label
changes the payoff of every incident parity constraint. In a binary
disjunction, the flip matters only if the other literal is false.

Earlier work showed that bounded degree gives a gain over Gaussian threshold
rounding for binary Boolean CSPs whose supplied SDP solution is
\emph{reasonable}~\cite[Theorem~23]{GHLM26}. Reasonableness bounds the SDP
biases and product moment on each edge and requires the score correlations of
incident edges to agree in sign after orientation by their predicates
(\cref{def:reasonable}). Suppose a fixed bounded function of the SDP bias gives
the rounding threshold at each variable, and the resulting assignment satisfies
an expected fraction $a$ of the total weight. On a reasonable solution, the
algorithm of that work achieves an expected fraction of at least
$a+\Omega(1/(d^4(\log d)^{7/2}))$. For Max 2-SAT, the same work uses the
geometry of the known configurations where LLZ rounding is tight to obtain the
approximation ratio $\alpha_{\rm LLZ}+\Omega(1/(d^2\log d))$.

Our first result gives an inverse-square gain for any Boolean 2-CSP language
that is \emph{regular near equality} at its baseline approximation ratio
$\alpha$ (\cref{def:near-equality}). The geometric requirement concerns the
binary constraints whose expected rounding payoff is close to $\alpha$ times
their SDP payoff. They must jointly have regular boundary geometry
(\cref{def:boundary}) for every Gaussian rule in a designated set. The baseline
must select a rule from this set with probability bounded below by a positive
constant. By \cref{lem:geometry}, jointly reasonable constraints in the sense
of~\cite{GHLM26} meet this requirement when their thresholds are bounded.

\begin{theorem}[Approximation under regularity near equality]
\label{cor:language-ratio}
Suppose a Boolean 2-CSP language is regular near equality at $0<\alpha<1$
(\cref{def:near-equality}) for an efficiently implementable baseline. On
weighted instances of occurrence degree at most $d\ge1$, a randomized
polynomial-time algorithm attains an expected approximation ratio of
$\alpha+\Omega(1/d^2)$, where the constant depends only on $\alpha$ and the
regularity parameters.
\end{theorem}

The rounding argument also gives an additive guarantee for a fixed Gaussian
threshold rule applied to any feasible solution of the SDP in
\cref{fig:relaxation}. Normalize the total weight to one. Write $a$ for the
expected weight satisfied by this rule and $w_{\rm bin}$ for the total weight
of constraints that depend on both variables.

\begin{theorem}[Additive gain over Gaussian threshold rounding]
\label{thm:general-rounding}
Let a weighted Boolean 2-CSP have occurrence degree at most $d\ge1$.
Fix a feasible solution of the SDP in \cref{fig:relaxation} and a Gaussian
threshold rule of expected value $a$. If the binary constraints have
regular boundary geometry (\cref{def:boundary}), a randomized
polynomial-time algorithm returns an assignment of expected value at
least $a+cw_{\rm bin}/d^2$, where $c>0$ depends only on the geometry
parameters.
\end{theorem}

For a reasonable binary SDP solution rounded with a fixed bounded threshold
function of the bias, \cref{thm:general-rounding} improves the additive gain
of~\cite{GHLM26} from $\Omega(1/(d^4(\log d)^{7/2}))$ to $\Omega(1/d^2)$.
The theorem also allows the thresholds to vary by vertex. The geometric
condition is strictly weaker than reasonableness. For example, take two
triangles sharing one vertex, with equality constraints on one triangle and
disequality constraints on the other. The SDP moments of independent unbiased
labels have regular boundary geometry at zero thresholds, although this
instance admits no reasonable SDP solution (\cref{sec:boundary-geometry}).

We count real arithmetic operations and comparisons, with unit cost for a
square root, a Gaussian draw, or an evaluation of the standard normal
distribution function or its inverse. Once the weights sum to one, solving the
SDP to additive accuracy $c_0/d^2$ suffices for \cref{cor:language-ratio}
whenever $c_0>0$ is sufficiently small.

For weighted Max 2-SAT, the LLZ thresholds near the two tight configurations
allow an inverse-linear gain. The property that gives this larger gain can be
stated for general binary predicates. Fix one variable's score at its rounding
threshold. Each incident binary constraint then has a boundary in the
remaining Gaussian coordinates. Orient the boundary toward the side where
assigning $+1$ to the fixed variable becomes more favorable. The sign condition in \emph{aligned
thresholds} (\cref{def:aligned-thresholds}) requires that the oriented
thresholds at each variable have the same sign and magnitudes bounded away from
zero. The sign may vary from one variable to another. For a mixture of rules,
it suffices that the baseline selects an aligned rule with probability bounded
below by a positive constant.

The next theorem applies to any certified baseline. Write $w_C$ for
the weight of a constraint $C$, $s_C$ for its SDP payoff, and $\bar a_C$
for its expected payoff under the baseline, averaged over the choice
of rounding rule and the rounding itself. Write $\val(\vect{x})$ for
the total weight satisfied by $\vect{x}$.

\begin{restatable}[Gain from aligned thresholds]{reptheorem}{alignedThresholdTheorem}
\label[theorem]{thm:aligned-sweep}
Fix a feasible SDP solution for a weighted Boolean 2-CSP of occurrence degree
at most $d\ge1$. Suppose the baseline assigns total probability at least
$\theta>0$ to a designated set of Gaussian rules. Let $\mathcal G$ be a
collection of binary constraints that has aligned thresholds with the same
parameters for every rule in this set. Let $0\le b_C\le\bar a_C$ be a target
payoff for each constraint. If every binary constraint $C\notin\mathcal G$
satisfies $b_C\le s_C$ and $\bar a_C-b_C\ge\tau s_C$ for some fixed $\tau>0$, a
randomized update returns an assignment $\vect{x}$ with
\[
 \Ex{\val(\vect{x})}\ge\sum_C w_C b_C+
       \frac{c}{d}\sum_{C\ {\rm binary}} w_Cs_C\mper
\]
Alternatively, if $\bar a_C-b_C\ge\tau$ for every binary constraint
$C\notin\mathcal G$, the bound holds with $w_{\rm bin}$ in place of the last
sum. The constant $c>0$ depends only on $\theta$, $\tau$, and the alignment
parameters. The update runs in polynomial time if we can implement the
baseline, recognize its designated rules, and find $\mathcal G$ in polynomial
time.
\end{restatable}

For a feasible SDP solution of value $s$, take $b_C=\alpha s_C$ in
\cref{thm:aligned-sweep}, with $0<\alpha<1$. The better of the updated
assignment and an assignment maximizing the unary and constant terms has
expected value at least $(\alpha+c'/d)s$ for some $c'>0$. When the hypotheses
hold uniformly, choosing an interior solution with
$s\ge(1-\varepsilon/d)\opt$ for sufficiently small fixed $\varepsilon>0$
gives an expected approximation ratio of $\alpha+\Omega(1/d)$
(\cref{sec:aligned-updates}). For weighted Max 2-SAT, we select the clauses
in fixed neighborhoods of the two tight LLZ configurations. Their thresholds
align, and every omitted binary clause has a uniformly positive payoff margin.

\begin{theorem}[Weighted Max 2-SAT]
\label{thm:main}
For some absolute constant $c>0$, a randomized polynomial-time algorithm
returns an assignment of expected value at least $(\alpha_{\rm LLZ}+c/d)\opt$
on every weighted Max 2-SAT instance of occurrence degree at most $d\ge1$.
\end{theorem}

Restricting the allowed disjunctions can raise the worst-case ratio.
In the convention of~\cite{BHZ23}, Max $\{1,2\}$-Horn-SAT has language
$\mathcal L_{\rm H}=\{x\lor y,\neg x\lor y,x,\neg x\}$.
Complementing every variable gives the usual Horn convention of at
most one positive literal per clause. The language
$\mathcal L_+=\{x\lor y,x,\neg x\}$ allows only positive binary
disjunctions, together with unit clauses of either sign.

\begin{theorem}[Restricted disjunction languages]
\label{thm:restricted}
For weighted instances of occurrence degree at most $d\ge1$, randomized
polynomial-time algorithms attain expected approximation ratios
$\alpha_{\rm H}+\Omega(1/d)$ for $\mathcal L_{\rm H}$ and
$\alpha_++\Omega(1/d)$ for $\mathcal L_+$, where
$\alpha_{\rm H}\simeq0.94615981$ and $\alpha_+\simeq0.95397990$ are the
respective sharp worst-case ratios under UGC.
\end{theorem}

\subsection{Prior Work and Comparison of Guarantees}

For unweighted Max-Cut, local changes after hyperplane rounding gave a gain of
$\Omega(1/d^4)$~\cite{FKL02}. A later analysis improved the gain to
$\Omega(1/d^3)$~\cite[Theorem~3.2.5]{FlorenThesis}. For weighted Max-2-Lin(2),
the analysis in~\cite{HK23} gave a gain of $\Omega(1/(d^2\log d))$. For
weighted Max-Cut and Max-2-Lin(2), the best known guarantee is now
$\alpha_{\rm GW}+\Omega(1/(d\log d))$~\cite[Theorem~1.1]{BGHKLMM26}.

For a binary disjunction, the SDP configuration records the two literal biases
and their product moment. An extremal distribution on these configurations was
identified in~\cite{Austrin07}. The proof of the simplicity conjecture
in~\cite{BHZ23} established the sharp LLZ ratio. Only two interior binary
configurations attain equality. Outside fixed neighborhoods of these
configurations, LLZ rounding has a uniformly positive margin~\cite[Theorem~12
and Lemma~13]{GHLM26}. This margin allowed the bounded-degree analysis
in~\cite{GHLM26} to focus on clauses in the two neighborhoods. The same local
improvement argument applies to a reasonable binary SDP solution for any
fixed bounded threshold function of the bias.

\Cref{tab:guarantees} summarizes the gains for reasonable instances and the
three disjunction languages.

\begin{table}[htbp]
\centering\small
\setlength{\tabcolsep}{4pt}\renewcommand{\arraystretch}{1.28}
\begin{tabular}{@{}lcc@{}}
\hline
Setting & Guarantee in cited work & This paper\\ \hline
Reasonable binary SDP solutions
 & $a+\Omega(1/(d^4(\log d)^{7/2}))$ & $a+\Omega(d^{-2})$\\
Max 2-SAT
 & $\alpha_{\rm LLZ}+\Omega(1/(d^2\log d))$
 & $\alpha_{\rm LLZ}+\Omega(d^{-1})$\\
Max $\{1,2\}$-Horn-SAT
 & $\alpha_{\rm H}$ & $\alpha_{\rm H}+\Omega(d^{-1})$\\
Positive binary disjunctions with signed units
 & $\alpha_+$ & $\alpha_++\Omega(d^{-1})$\\ \hline
\end{tabular}
\NextLinkTarget{comparison-table}
\caption{Guarantees for weighted instances of occurrence degree $d\ge2$.
In the first row, the weights sum to one and $a$ is the expected weight
satisfied using a fixed bounded threshold function of the SDP bias. The other
rows give approximation ratios. The earlier guarantees in the first two rows
are from~\cite{GHLM26}, and those in the last two are from~\cite{BHZ23}.}
\label{tab:guarantees}
\end{table}

\subsection{Proof Overview}

\paragraph{Updating in reverse order.}
For the inverse-square gain, we reconsider a variable's label when its Gaussian
score is near its threshold. Two proposed improvements can interfere if both
endpoints of a constraint are in their windows. Consider a disequality whose
endpoints have the same label. Flipping either endpoint alone satisfies the
edge. Flipping both leaves it unsatisfied.

The reverse sweep avoids counting both gains against the original labels. A
forward pass sets each threshold shift using the provisional thresholds of
earlier neighbors. We then reveal scores and update labels from last to first.
When we visit a variable, the later neighbors already have final labels. For
the earlier neighbors we average over their provisional labels conditional on
the scores revealed so far. In the example, once the first endpoint flips, the
edge favors keeping the second endpoint's label. The changes in satisfied
weight telescope in sweep order to the gain of the final assignment.

A binary constraint's \emph{vote} at a variable is the difference between its
payoffs from assigning $+1$ and $-1$, given the other variable's label. For a
disjunction, the vote is zero when the other literal is true. Otherwise it
favors the label that satisfies the clause. A weighted sum of these votes tells
us which label to choose. To estimate the available gain at a variable, we
freeze its score at its threshold, keep the orthogonal part of the Gaussian
fixed, and recompute its neighbors' baseline labels. The resulting vote from
the selected binary constraints and the unit constraints is independent of the
variable's own score. If the frozen vote is nonzero, the baseline label
disagrees with it on one half of the threshold window. When the threshold is
bounded and the window has half-width $h\le1$, that half has Gaussian
probability $\Omega(h)$.

\paragraph{How much of the vote is visible?}
For the reverse sweep we estimate the part of the frozen vote that can be
predicted from the revealed scores. With the score at $i$ frozen, each selected
constraint $C=ij$ contributes a constant plus a weighted halfspace indicator in
the remaining Gaussian coordinates. Write $\vect{e}_{iC}$ for its unit normal,
oriented so that entering the halfspace increases the vote. The parity analysis
of~\cite{BGHKLMM26} uses these projected directions for pairs of edges sharing
one endpoint. Here the angle between two such normals controls how often the
corresponding votes change in opposite directions. We perturb the revealed
coordinates in opposite directions by an amount of order $\lambda$ and complete
both draws with the same hidden coordinates. If the normal of a halfspace has
been revealed, the two draws fall on opposite sides of its boundary with
probability $\Omega(\lambda)$. Crossings in opposite directions can cancel.
Regular boundary geometry bounds the probability of such a pair by
$O(\lambda^2)$, and each constraint has at most $d-1$ partners at either
endpoint. When the second endpoint is visited, the first has already been
revealed. Summing over the sweep and taking $\lambda$ to be a small multiple of
$1/d$ gives a total expected absolute visible vote of $\Omega(w/d)$ for
selected weight $w$ (\cref{lem:crossings}). A frozen vote can overstate the
actual gain on a selected constraint only when both endpoint scores are within
$O(h)$ of their thresholds. Their bounded joint density gives probability
$O(h^2)$ for this event (\cref{lem:payoff}). Thus choosing $h$ a sufficiently
small multiple of $1/d$ gives the inverse-square gain.

\paragraph{Why aligned thresholds give a larger gain.}
Consider one of the Gaussian rules on which the selected constraints have
aligned thresholds (\cref{def:aligned-thresholds}). Fix a variable $i$
incident to a selected constraint. By negating its vote and complementing
its indicators if necessary, we may take its oriented thresholds to be
positive. The vote depends only on the span of its selected normals.
Let $k$ be its dimension, so $1\le k\le d$. Stretch the Gaussian in this
span by $1+1/\sqrt{k}$. Every crossed halfspace is entered. Write
$\phi(x)=(2\pi)^{-1/2}\exp(-x^2/2)$ for the standard normal density. At
threshold $\tau$, the interval of scores that cross has Gaussian mass
$\Omega(\tau\phi(\tau)/\sqrt{k})$. A constant fraction of each crossing
occurs at radii where the density of the stretched point is at most a
constant multiple of the original Gaussian density. Let $w_i$ be the sum
of the selected halfspace weights at $i$. Since alignment bounds
$\tau\phi(\tau)$ away from zero, \cref{lem:sat-quadratic-vote} gives an
expected absolute frozen vote of $\Omega(w_i/\sqrt d)$. We compute all
frozen votes from the same Gaussian draw, activate each variable independently
with a small constant probability, and update active variables whose scores lie
in their windows. The predicted gain per unit of selected weight is now
$\Omega(h/\sqrt d)$. An error on a selected constraint still requires both
endpoint scores to be near their thresholds and costs $O(h^2)$ per unit weight.
A window of half-width $h=\kappa/\sqrt d$ for sufficiently small $\kappa>0$
gives the inverse-linear gain (\cref{thm:aligned-sweep}).

For Max 2-SAT, the common sign can be read off at either tight configuration.
The two literal thresholds have the same nonzero sign and the literal scores
are negatively correlated. Consequently every near-tight clause at a variable
has an oriented threshold of sign opposite to that variable's rounding
threshold, whichever polarity occurs in the clause (\cref{sec:sat}). In all
three disjunction languages the remaining binary clauses have a uniform
positive margin over their target payoff. The theorem can also use a margin
proportional to the SDP payoff of an omitted constraint. With probability
$(1-p)^2$ neither endpoint is activated, where $p$ is the activation
probability. On that event the constraint keeps its baseline payoff. Choosing
$p$ small enough preserves a fixed fraction of the margin, even for constraints
with arbitrarily small SDP payoff.

\subsection{Comparison of Proof Techniques}

The Max-Cut algorithms of~\cite{FKL02,FlorenThesis} look for vertices whose
flip improves the rounded cut. The parity algorithm of~\cite{HK23} marks
vertices near the rounding hyperplane and flips one when the preference of its
unmarked neighbors outweighs all marked neighbors. A different parity algorithm
shifts all thresholds simultaneously using smoothed votes. Its analysis expands
the correlation of the smoothed and unsmoothed votes at $i$ as a sum over pairs
of incident edges. For distinct neighbors $j,k$, the summand depends on the
inner product of the signed projections for $ij$ and $ik$~\cite[Algorithm~1 and
Lemma~3.10]{BGHKLMM26}. For Boolean 2-CSPs, the weighted-count update
of~\cite[Algorithm~1]{GHLM26} assigns a score to every neighborhood. It tests
whether the net vote from neighbors whose labels are fixed is enough to cover
the total unresolved weight. For near-tight Max 2-SAT clauses, its analysis
bounds the conditional variance to obtain an expected absolute vote of
$\Omega(w_i/d)$ at a variable of incident weight $w_i$. The update uses a
window of order $1/(d\log d)$~\cite[Theorems~14--15 and Lemma~16]{GHLM26}.

Both of our estimates address the expected absolute vote directly. For the
reverse sweep, the bound on opposite crossings uses the angle between oriented
normals and allows thresholds of either sign, including zero, as in the
two-triangle example of \cref{sec:boundary-geometry}. For the inverse-linear
gain, the common sign prevents cancellation along a ray, and uniform
separation from zero makes each selected
halfspace contribute at least a constant times its weight divided by $\sqrt d$
in \cref{eq:aligned-radial}. The radial comparison permits arbitrary angles
between incident normals and an arbitrary constant contribution from unit
clauses, including units on which the baseline guarantee is tight.

Earlier Max-Cut analyses distinguish edges near the critical correlation from
the rest~\cite{FKL02,FlorenThesis,HK23}. For Max 2-SAT, the proof
of~\cite[Theorem~15]{GHLM26} uses LLZ rounding if the fraction of clause weight
outside the near-tight set reaches a threshold of order $1/(d^2\log d)$, and
otherwise applies the local update to the near-tight clauses. The parity
algorithm of~\cite[Algorithm~1]{BGHKLMM26} compares the fraction of SDP value
outside the tight interval with a threshold of order $1/(d\log d)$. Our reverse
sweep runs on the full instance. The aligned update uses the selected
constraints to form its binary votes. In both algorithms, the choice of window
width is independent of the total weight of near-tight constraints. The aligned
update also fixes its activation probability independently of that weight. The
baseline slack on the remaining constraints covers their expected loss
(\cref{lem:binary-weight,sec:aligned-updates}).

\paragraph{Organization of the paper.}
\Cref{sec:prelim} gives the SDP relaxation and defines regular boundary
geometry and aligned thresholds. \Cref{sec:gain} analyzes the reverse sweep and
proves the two inverse-square guarantees. \Cref{sec:linear-gain} proves the
inverse-linear gain in \cref{thm:aligned-sweep}. \Cref{sec:sat} verifies
alignment for Max 2-SAT and the two restricted disjunction languages.

\section{Preliminaries and Notation}\label{sec:prelim}
We use ordinary lowercase letters for scalars, bold lowercase letters for
vectors, and uppercase letters for matrices. Scalar scores and votes keep the
same notation before and after a Gaussian draw. We also use nonbold uppercase
letters for random quantities such as the frozen vote $D_i$.

\subsection{Instances and Notation}

Let $[n]=\{1,\ldots,n\}$. An instance is a finite collection $\mathcal C$ of
constraints on $n$ variables with labels in $\{\pm 1\}$. Constraint $C$ has an
ordered scope of at most two distinct variables, a predicate $f_C$ mapping
their labels to $\{0,1\}$, and a positive weight $w_C$. Remove any input on
which the predicate does not depend. Thus a binary constraint in what follows
depends on both inputs. We write $i\in C$ when $i$ belongs to the scope of $C$,
and $C=ij$ for a binary constraint with endpoints $i,j$. In a sum at fixed $i$,
$C=ij$ ranges over all incident binary constraints, with $j$ the other
endpoint. Parallel constraints count separately. The occurrence degree of a
variable is the number of constraints containing it, and $d\ge1$ is an upper
bound on these degrees. We normalize the weights so that
$\sum_{C\in\mathcal C}w_C=1$, scaling every assignment value and the optimum by
the same factor.

For an assignment $\vect{x}=(x_1,\ldots,x_n)\in\{\pm 1\}^n$, write
$\vect{x}_C$ for its restriction to the ordered scope of $C$.
The value of the assignment and the optimum are
\[
   \val(\vect{x})=\sum_{C\in\mathcal C}w_C f_C(\vect{x}_C),\qquad
   \opt=\max_{\vect{x}\in\{\pm 1\}^n}\val(\vect{x})\mper
\]
We use $\innerprod{\vect{u}}{\vect{v}}$ for the Euclidean inner product and
$\norm{\vect{u}}=\sqrt{\innerprod{\vect{u}}{\vect{u}}}$ for its norm. The notation
$\Span\{\vect{v}_1,\ldots,\vect{v}_k\}$ denotes the linear span of the listed vectors,
and $L^\perp$ denotes the orthogonal complement of a subspace $L$.
We write $\Ex{\cdot}$ and
$\Pr{\cdot}$ for expectation and probability, $\1{E}$ for the indicator of
an event $E$, and $\operatorname{sgn}(x)$ for the sign of a nonzero real $x$.
For symmetric matrices, $A\succeq B$ means $A-B$ is positive semidefinite.

\subsection{The Semidefinite Relaxation}

\Cref{fig:relaxation} gives the relaxation we round~\cite[Section~2.1]{GHLM26}.
For a feasible Gram matrix $K=(k_{ij})$, write
$k_{ij}=\innerprod{\vect{v}_i}{\vect{v}_j}$. Its diagonal entries are
one, so $\vect{v}_0,\ldots,\vect{v}_n$ are unit vectors. We take
$\vect{v}_0$ as the reference vector and write $\xi_i=k_{0i}$ for the bias of variable
$i$ and $\rho_{ij}=k_{ij}$ for the product moment of variables $i,j$.

\begin{figure}[htbp]
\centering
\begin{minipage}{0.95\linewidth}
\small
The decision variable $K=(k_{ij})$ is a real symmetric matrix indexed by
$\mathcal I=\{0,1,\ldots,n\}$. For signs $a,b\in\{\pm 1\}$, the local probabilities for the empty, unary,
and binary scopes are
\[
 p_{\varnothing}(\varnothing)=1,\qquad
 p_i(a)=\frac{1+ak_{0i}}{2},\qquad
 p_{ij}(a,b)=\frac{1+ak_{0i}+bk_{0j}+abk_{ij}}{4}\mper
\]
For a constraint $C$ with ordered scope $\mathcal S$, its SDP payoff is
\[
 s_C=\sum_{\vect{x}\in\{\pm 1\}^{|\mathcal S|}}
             f_C(\vect{x})\,p_{\mathcal S}(\vect{x})\mper
\]
\[
\begin{alignedat}{2}
 \text{maximize}\quad & s=\sum_{C\in\mathcal C}w_C s_C \\
 \text{subject to}\quad & K\succeq 0, \\
 & k_{ii}=1 &&\qquad (i\in \mathcal I),\\
 & 1+\epsilon \delta k_{ij}-\epsilon \zeta k_{ik}-\delta \zeta k_{jk}\ge0
       &&\qquad (i,j,k\in \mathcal I,\ \epsilon,\delta,\zeta\in\{\pm 1\})\mper
\end{alignedat}
\]
\end{minipage}
\caption{The SDP relaxation for weighted Boolean 2-CSPs with all signed
triangle inequalities.}\label{fig:relaxation}
\end{figure}

The last constraint in \cref{fig:relaxation} is the $\ell_2^2$-triangle inequality
$\innerprod{\epsilon\vect{v}_i-\zeta\vect{v}_k}{\delta\vect{v}_j-\zeta\vect{v}_k}\ge0$
at $\zeta\vect{v}_k$. \Cref{lem:projection} uses these inequalities to bound
the angle between projected neighbor vectors in a reasonable collection.
Taking $k=0$ and $(\epsilon,\delta,\zeta)=(a,b,-1)$ gives $4p_{ij}(a,b)\ge0$.
Summing over either sign gives the marginals $p_i,p_j$. Thus $p_{ij}$ is a
distribution with biases $\xi_i,\xi_j$ and product moment $\rho_{ij}$.

Every assignment $\vect{x}\in\{\pm 1\}^n$ gives feasible vectors
$\vect{v}_i=x_i\vect{v}_0$. All signed vectors lie in
$\{\vect{v}_0,-\vect{v}_0\}$ and satisfy the triangle inequalities. For each
constraint $C$, the local distribution puts unit mass on $\vect{x}_C$, giving
$s=\val(\vect{x})$. It follows that the optimal SDP value is at least $\opt$.
It is at most one because $s_C\in[0,1]$ and the weights sum to one.

\subsection{Gaussian Threshold Rounding}

We call the bias $\xi_i$ \emph{interior} if $|\xi_i|<1$. In this case let
$\vect{u}_i=(\vect{v}_i-\xi_i \vect{v}_0)/\sqrt{1-\xi_i^2}$, the normalized
projection of $\vect{v}_i$ onto $\vect{v}_0^\perp$. For each $i$ with
$|\xi_i|=1$, choose a new unit vector $\vect{u}_i$ orthogonal to all SDP
vectors and to the other new directions, enlarging the space if necessary. Draw
a standard Gaussian vector $\vect{g}$ in this space. Its coordinates in any
orthonormal basis are independent standard normals. The scores
$z_i=\innerprod{\vect{u}_i}{\vect{g}}$ are standard normal with correlations
$r_{ij}=\innerprod{\vect{u}_i}{\vect{u}_j}$, by linearity and the identity
covariance of $\vect{g}$. When both biases are interior, substituting the
normalized projections into $\innerprod{\vect{u}_i}{\vect{u}_j}$ gives
\[
  r_{ij}=\frac{\rho_{ij}-\xi_i\xi_j}{\sqrt{(1-\xi_i^2)(1-\xi_j^2)}}\mper
\]

For Max-Cut and Max-2-Lin(2), we may take every bias to be zero. Average
$K$ with the Gram matrix obtained by negating every variable vector while
fixing $\vect{v}_0$. Both matrices satisfy the signed triangle inequalities,
and their average is feasible. The biases cancel and all edge moments, hence
the SDP value, are preserved. In this solution $\vect{u}_i=\vect{v}_i$ and
zero thresholds give the usual hyperplane rounding~\cite{GW95}.

For real $t,z$, write $\thr_t(z)=-1$ if $z\le t$ and $+1$ otherwise.
Before drawing $\vect{g}$, a rounding rule specifies either a real threshold
$t_i$ or a deterministic label at each variable $i$. The baseline assignment
sets $y_i^0=\thr_{t_i}(z_i)$ in the first case and uses the specified label in
the second. Define the baseline payoff $a_C=\Ex{f_C(\vect{y}_C^0)}$, so that
$\Ex{\val(\vect{y}^0)}=\sum_Cw_Ca_C$ by linearity. Write
$\phi(z)=(2\pi)^{-1/2}\exp(-z^2/2)$ for the standard normal density and
$\Phi(z)=\int_{-\infty}^{z}\phi(x)\,dx$ for its distribution function. At a
finite threshold, $\Ex{y_i^0}=1-2\Phi(t_i)$, so a rounded bias
$f(\xi_i)\in(-1,1)$ gives $t_i=\Phi^{-1}((1-f(\xi_i))/2)$. A rounded bias of
$+1$ or $-1$ gives that label deterministically. Both updates keep deterministic
labels fixed and form shifts, windows, and frozen scores only at variables
with finite thresholds. An upper-tail threshold for label $-1$
in~\cite{GHLM26} corresponds to its negative here, since $\vect{g}$ and
$-\vect{g}$ have the same law.

When the baseline chooses among rounding rules before drawing $\vect{g}$, write
$\bar a_C$ for the payoff averaged over both the choice of rule and the
randomness of that rule. For a fixed Gaussian rule, $\bar a_C=a_C$. Relative to
a comparison ratio $\alpha$, the \emph{slack} of $C$ is
$\ell_C=\bar a_C-\alpha s_C$. A constraint is tight if its slack is zero.

\subsection{Reasonableness, Boundary Geometry, and Aligned Thresholds}
\label{sec:boundary-geometry}

Abbreviate $+1$ and $-1$ by $+$ and $-$ in truth tables. For a binary predicate put
\[
 m_C=f_C(+,+)-f_C(+,-)-f_C(-,+)+f_C(-,-)\mper
\]
Interchanging the endpoints preserves this mixed difference. To define the vote
of $C$ at $i$, write $i$ first, transposing the truth table if necessary. If
the other endpoint has label $y$, the vote is
$\delta_{iC}(y)=f_C(+,y)-f_C(-,y)$. Changing $y$ from $-1$ to $+1$ adds $m_C$
to the vote, so
\[
 \delta_{iC}(y)=\delta_{iC}(-1)+m_C\1{y=+1}\mper
\]
For a unary predicate, the vote is the constant $f_C(+)-f_C(-)$. In every case
its magnitude is at most one because it is a difference of two bits. Four
binary predicates accept exactly one label pair, and four reject exactly one.
Of the six truth tables accepting two pairs, four depend on only one variable
and the other two are parity predicates. Thus $|m_C|=1$ for the first eight
predicates and $|m_C|=2$ for parity.

In the local SDP distribution, $\xi_i$ is the mean label at $i$ and
$\rho_{ij}$ is the mean product of the labels at $i$ and $j$. The
Gaussian scores used for rounding have correlation $r_{ij}$.

\newtheorem{cspdefinition}{Definition}[section]
\def\theHcspdefinition{\theHsection.\arabic{cspdefinition}}
\begin{cspdefinition}[Reasonableness]\label[definition]{def:reasonable}
A subcollection $\mathcal G\subseteq\mathcal C$ is reasonable for the
supplied SDP vectors if each $C=ij\in\mathcal G$ depends on both inputs,
\[
 |\xi_i|,|\xi_j|\le 0.4,\qquad 0.5\le|\rho_{ij}|\le0.9\mcom
\]
and, at each vertex $i$, the quantities $m_Cr_{ij}$ have a common sign
over the incident constraints $C=ij\in\mathcal G$.
\end{cspdefinition}

This is the alignment condition in Definitions~19 and~20 of~\cite{GHLM26}.
Indeed, for the ten truth tables just enumerated,
$y\delta_{iC}(y)=\operatorname{sgn}(m_C)$ whenever the vote is nonzero. If only
one vote is nonzero, this identity follows directly from the difference
formula. If both are nonzero, they have opposite signs. For labels $y,z$ at
which the respective votes are nonzero, the alignment test compares
$\delta_{iC}(y)$ with $yz\delta_{iC'}(z)$. Multiplying both by $y$ gives the
signs of the two mixed differences. Thus aligned predicates have mixed
differences of the same sign, and anti-aligned predicates have mixed
differences of opposite signs. The resulting condition on the score
correlations is exactly the one in \cref{def:reasonable}. The score
correlations on $\mathcal G$ are nonzero since
$|\xi_i\xi_j|\le0.16<0.5\le|\rho_{ij}|$.

For any binary constraint $C=ij$ with $|r_{ij}|<1$, projecting
$\vect{u}_j$ onto $\vect{u}_i^\perp$ gives $\vect{u}_j-r_{ij}\vect{u}_i$.
Changing the neighbor's label from $-1$ to $+1$ changes the vote at $i$ by
$m_C$. We orient the projected unit normal toward the side on which this
vote increases, giving
\[
 \vect{e}_{iC}=\operatorname{sgn}(m_C)
       \frac{\vect{u}_j-r_{ij}\vect{u}_i}{\sqrt{1-r_{ij}^2}}\mper
\]

\begin{cspdefinition}[Regular boundary geometry]\label[definition]{def:boundary}
Fix a feasible SDP solution and a Gaussian threshold rule. A collection
$\mathcal G$ of binary constraints has \emph{regular boundary geometry}
with parameters $t_*\ge0$, $0\le r_*<1$ and $0<\gamma\le1$ if
\begin{enumerate}[label=\textup{(\roman*)},leftmargin=*]
\item At every $C=ij\in\mathcal G$,
$|t_i|,|t_j|\le t_*$ and $|r_{ij}|\le r_*$.
\item At each $i$, any two distinct incident constraints $C,C'\in\mathcal G$,
including parallel constraints, satisfy
$\innerprod{\vect{e}_{iC}}{\vect{e}_{iC'}}\ge-1+\gamma$.
\end{enumerate}
\end{cspdefinition}

In the zero-bias SDP for parity, an edge $C=ij$ requiring $x_ix_j=b_{ij}$,
with $b_{ij}\in\{\pm1\}$, has mixed difference $m_C=2b_{ij}$. Hence
$\innerprod{\vect{g}}{\vect{e}_{iC}}$ is exactly the Gaussian coordinate
$h_{ij}$ used in~\cite[Algorithm~1]{BGHKLMM26}. For edges $ij$ and $ik$ sharing
only $i$, the correlation of $h_{ij}$ and $h_{ik}$ is the inner product of
their normals. For general binary predicates, the same inner product can be
expressed in terms of the score correlations, including the correlation
$r_{jk}$ between the two neighbors. Expanding the projected normals gives
\[
 \innerprod{\vect{e}_{iC}}{\vect{e}_{iC'}}=
 \operatorname{sgn}(m_C m_{C'})
 \frac{r_{jk}-r_{ij}r_{ik}}{\sqrt{(1-r_{ij}^2)(1-r_{ik}^2)}}\mper
\]
Call the members of the chosen binary collection $\mathcal G$
\emph{good}, and put $\widehat w_C=w_C|m_C|$ for each of them.
At a vertex $i$ with a finite threshold, \emph{freezing its score} means replacing
$z_i$ by $t_i$ and keeping $\vect{g}_i^\perp=\vect{g}-z_i \vect{u}_i$ fixed. The resulting
score for $j$ is
\[
 z_j^{[i]}=\innerprod{\vect{u}_j}{\vect{g}_i^\perp}+r_{ij}t_i
           =z_j-r_{ij}(z_i-t_i)\mper
\]
The first expression uses $\vect{g}_i^\perp+t_i \vect{u}_i$ in place
of $\vect{g}$, and the second uses $\innerprod{\vect{u}_j}{\vect{u}_i}=r_{ij}$.
The \emph{frozen vote} $D_i$ at $i$ includes every unary constraint at
$i$ and every incident binary constraint in $\mathcal G$. It equals
\[
 D_i=\sum_{\substack{C\text{ unary}\\i\in C}} w_C\bigl(f_C(+)-f_C(-)\bigr)
      +\sum_{C=ij\in\mathcal G}w_C\delta_{iC}(\thr_{t_j}(z_j^{[i]}))\mper
\]
In an orthonormal basis starting with $\vect{u}_i$, the first coordinate of
$\vect{g}$ is $z_i$. The remaining coordinates determine $\vect{g}_i^\perp$ and
are independent of $z_i$. The frozen vote $D_i$ depends only on these remaining
coordinates and is therefore independent of $z_i$.

When we freeze $z_i$, the resulting score at the other endpoint $j$ has mean
$r_{ij}t_i$ and standard deviation $\sqrt{1-r_{ij}^2}$. We orient each
constraint's halfspace toward the side on which its vote at $i$ increases. This
lets us compare thresholds at a shared variable.

\begin{cspdefinition}[Aligned thresholds]
\label[definition]{def:aligned-thresholds}
Fix a feasible SDP solution and a Gaussian threshold rule. For a binary
constraint $C=ij$ with $|r_{ij}|<1$, its \emph{oriented threshold}
at $i$ is
\[
 \tau_{iC}=\operatorname{sgn}(m_C)
              \frac{t_j-r_{ij}t_i}{\sqrt{1-r_{ij}^2}}\mper
\]
A collection $\mathcal G$ of binary constraints has \emph{aligned thresholds} with
parameters $0\le t_*<\infty$, $0\le r_*<1$, and $0<\tau_0\le\tau_*<\infty$ if
\begin{enumerate}[label=\textup{(\roman*)},leftmargin=*]
\item Each $C=ij\in\mathcal G$ satisfies
$|t_i|,|t_j|\le t_*$ and $|r_{ij}|\le r_*$.
\item For each variable $i$ there is a sign $\epsilon_i\in\{\pm1\}$
such that $\tau_0\le\epsilon_i\tau_{iC}\le\tau_*$ for every incident
$C\in\mathcal G$.
\end{enumerate}
\end{cspdefinition}

The sign is common to the selected constraints at each variable. It may vary
between variables, and the incident normals may form arbitrary angles.

Sampling a constraint from a nonempty reasonable collection in proportion to
its weight gives a reasonable distribution of configurations in the terminology
of~\cite{GHLM26}. Write $\vect{v}_j^{\perp\{0,i\}}$ for the normalized
projection of $\vect{v}_j$ onto $\Span\{\vect{v}_0,\vect{v}_i\}^{\perp}$
whenever the projection is nonzero. The proof of \cref{lem:geometry} shows that
this projection is nonzero on every reasonable edge. At a shared endpoint, the
predicates of $C$ and $C'$ are aligned if their mixed differences $m_C$ and
$m_{C'}$ have the same sign and anti-aligned if the signs are opposite.

\begin{lemma}[Lemma~22~\cite{GHLM26}]
\label{lem:projection}
Suppose the SDP vectors satisfy the signed triangle inequalities. Let $C=ij$
and $C'=ik$ be constraints in the same reasonable subcollection, with the
common endpoint $i$ written first in both predicates. The normalized
projections satisfy
\[
 \begin{aligned}
 \innerprod{\vect{v}_j^{\perp\{0,i\}}}{\vect{v}_k^{\perp\{0,i\}}}&\ge-5/6
     &&\text{if the predicates are aligned},\\
 \innerprod{\vect{v}_j^{\perp\{0,i\}}}{\vect{v}_k^{\perp\{0,i\}}}&\le 5/6
     &&\text{if the predicates are anti-aligned}\mper
 \end{aligned}
\]
\end{lemma}

The frozen vote $D_i$ sums the weighted unary and selected binary votes at $i$
with $z_i=t_i$ and the orthogonal component
$\vect{g}_i^\perp=\vect{g}-z_i\vect{u}_i$ held fixed. Crossing the boundary of
$C$ along its oriented unit normal $\vect{e}_{iC}$ increases the vote by
$\widehat w_C=w_C|m_C|$. In \cref{def:boundary}, $t_*$ bounds the threshold
magnitudes and $r_*$ bounds the absolute score correlations. The normals of
distinct constraints incident to the same variable have inner product at least
$-1+\gamma$.

\begin{lemma}[Geometry at a threshold]\label{lem:geometry}
For a collection with regular boundary geometry, $D_i$ is a constant plus a
weighted sum of halfspace indicators evaluated at $\vect{g}_i^\perp$. These
halfspaces have weights $\widehat w_C$, normals $\vect{e}_{iC}$, and thresholds
of absolute value at most $2t_*/\sqrt{1-r_*^2}$. Every reasonable collection
with endpoint thresholds bounded by $t_*$ has this geometry with $\gamma=1/6$
and an absolute $r_*<1$.
\end{lemma}

\begin{proof}
First consider a reasonable collection. Both labels on a reasonable edge have
positive variance under the local SDP distribution. If $|r_{ij}|=1$, equality
in Cauchy--Schwarz for the centered labels forces one label to be an affine
function of the other almost surely. Each label takes both values in
$\{\pm 1\}$, so this function is either the identity or negation. Either case
gives $|\rho_{ij}|=1$, contradicting \cref{def:reasonable}. The local
distributions satisfying the bounds on the biases and the product moment form a
compact subset of the four-point probability simplex. The score correlation is
continuous there because both variances are bounded away from zero. Hence its
absolute value has a uniform upper bound $r_*<1$.

For any binary constraint with $|r_{ij}|<1$, write $s=\operatorname{sgn}(m_C)$
and $r=r_{ij}$. Its oriented halfspace has threshold $\tau_{iC}$ as given by
\cref{def:aligned-thresholds}. The residual $\vect{u}_j-r \vect{u}_i$ is
orthogonal to $\vect{u}_i$ and has squared norm $1-r^2$. For $s=+1$, the
variable part of the vote equals
$\widehat w_C\1{\innerprod{\vect{e}_{iC}}{\vect{g}_i^\perp}\ge \tau_{iC}}$ by
the vote formula. For $s=-1$ use $-\1{H}=\1{H^c}-1$ and reverse the normal and
threshold. In both cases the boundary has probability zero because $1-r^2>0$.
Under \cref{def:boundary}(i) this gives $|\tau_{iC}|\le2t_*/\sqrt{1-r_*^2}$. It
remains to check the angles for a reasonable collection. The unsigned residual
normal in $\vect{e}_{iC}$ is $\vect{v}_j^{\perp\{0,i\}}$. Indeed, projecting
$\vect{v}_j=\xi_j\vect{v}_0+\sqrt{1-\xi_j^2}\,\vect{u}_j$ onto
$\Span\{\vect{v}_0,\vect{v}_i\}^{\perp}$ leaves
$\sqrt{1-\xi_j^2}(\vect{u}_j-r \vect{u}_i)$. For distinct neighbors, the mixed
differences have the same sign for aligned predicates and opposite signs for
anti-aligned predicates. By \cref{lem:projection}, applied with the signed
$\ell_2^2$-triangle inequalities of \cref{fig:relaxation}, the oriented normals
have inner product at least $-5/6$ in both cases. Parallel constraints have the
same nonzero score correlation $r_{ij}$. By \cref{def:reasonable}, their mixed
differences therefore have the same sign, so their oriented normals coincide.
\end{proof}

To see that regular boundary geometry strictly extends reasonableness,
consider two triangles sharing exactly one vertex. Put equality constraints
on one triangle and disequality constraints on the other. Independent unbiased
labels give the identity Gram matrix. With thresholds zero, the normal at
$i$ for an incident edge $ij$ is either $\vect{u}_j$ or $-\vect{u}_j$.
Since distinct neighbors have orthogonal score directions,
\cref{def:boundary} holds with $(t_*,r_*,\gamma)=(0,0,1)$.

Suppose this instance admitted a reasonable SDP solution. On every edge,
$|\rho_{ij}|\ge1/2>|\xi_i\xi_j|$, so $r_{ij}$ has the sign of $\rho_{ij}$.
By \cref{def:reasonable}, the sign of $m_Cr_{ij}$ agrees on incident edges,
and hence on the whole connected graph. Since $m_C=2$ on the equality
triangle and $m_C=-2$ on the disequality triangle, one triangle would have
all three product moments at most $-1/2$. Its vertices $i,j,k$ would then
violate the signed triangle inequality
$1+\rho_{ij}+\rho_{ik}+\rho_{jk}\ge0$ in \cref{fig:relaxation}.

\subsection{Rounding Rules and Tight Configurations for Max 2-SAT}\label{sec:clause-baselines}

Represent true by $-1$. Within a clause, a literal on variable $i$ has value
$\epsilon_i x_i$, where its sign $\epsilon_i\in\{\pm 1\}$ is $+1$ for an
unnegated variable and $-1$ for a negated one. For two literal values $x,y$,
the disjunction has payoff
\[
 f_{\lor}(x,y)=1-\frac{(1+x)(1+y)}4=\frac{3-x-y-xy}{4}\mper
\]
The clause fails only when both literals are $+1$. A clause containing the same
literal twice is unary, and a clause containing complementary literals is a
constant of payoff one. These simplifications do not increase the occurrence
degree. An empty clause, if allowed, always contributes zero and can be
discarded before normalization without changing any ratio.

The SDP moments in literal coordinates for $C=ij$ are
\[
  a=\epsilon_i\xi_i,\qquad b=\epsilon_j\xi_j,\qquad
  \rho=\epsilon_i\epsilon_j\rho_{ij}\mper
\]
Thus $a,b$ are the literal biases and $\rho$ is their product moment.
The payoff formula gives $s_C=(3-a-b-\rho)/4$.

\paragraph{The LLZ rule.}
Set $\beta=\alpha_{\rm LLZ}$. The LLZ rule~\cite{LewinLZ02,BHZ23} has rounded
bias $f(\xi)=\beta\xi$, so its threshold at bias $\xi$ is
$t(\xi)=\Phi^{-1}((1-\beta\xi)/2)$. Since $0<\beta<1$, monotonicity of $\Phi$
and the identity $\Phi(-z)=1-\Phi(z)$ give the uniform bound
$|t(\xi)|\le\Phi^{-1}((1+\beta)/2)$. The same symmetry makes $t$ odd. Thus
negating a literal negates its baseline label under the corresponding signed
score, except on the probability-zero event of equality at the threshold.

For a binary clause with interior endpoint biases, the LLZ analysis gives
$\ell_C=a_C-\beta s_C\ge0$, with equality
exactly at two triples
\[
 (a,b,\rho)=(u,u,-1+2u)\quad\text{or}\quad(-u,-u,-1+2u),\qquad
 u\simeq0.162478\mper
\]
Outside any fixed neighborhoods of these two triples, $\ell_C$ has a uniform
positive lower bound. Theorem~12 and Lemma~13 of~\cite{GHLM26} give a quadratic
separation bound using the tightness analysis in~\cite{BHZ23}. The bound
applies with our sign convention because negating a literal negates its score,
and replacing $\vect{g}$ by $-\vect{g}$ interchanges the lower- and upper-tail
conventions.

\paragraph{The two restricted languages.}
For the Horn language $\mathcal L_{\rm H}$ and the positive binary
language $\mathcal L_+$ of \cref{sec:intro}, use the respective ratios
$\alpha=\alpha_{\rm H}$ and $\alpha=\alpha_+$. The following rounding
rules satisfy $\ell_C=\bar a_C-\alpha s_C\ge0$ for every binary
clause~\cite[full version, Proposition~4.1 and Lemma~5.2]{BHZ23}.

\begin{itemize}[leftmargin=*]
\item For $\mathcal L_{\rm H}$, use the Gaussian rule $f(b)=b$ with
probability $\alpha_{\rm H}$ and set every variable to true otherwise.
Every allowed binary clause contains a positive literal, so the all-true
assignment satisfies it.
\item For $\mathcal L_+$, always use the Gaussian rule
$f(b)=-1+\alpha_+(1+b)$. A variable of bias $b$ therefore rounds
to false with probability $\alpha_+(1+b)/2$.
\end{itemize}

Under the Horn Gaussian rule, $f(b)=b$ is odd, so negating a literal negates
its rounded label under the signed score. The same formula for the probability
of satisfying a disjunction therefore applies to every allowed choice of
literal signs. For $\mathcal L_+$ both literals are positive. The proofs of
Theorems~4.4 and~5.5 in the full version of~\cite{BHZ23} locate the binary
configurations of minimum slack. In both languages every such configuration
satisfies $\rho=-1+|a+b|$. For $\mathcal L_+$, $a,b$ are both within $10^{-6}$
of $-0.1824167935$. For $\mathcal L_{\rm H}$ they are both within $10^{-6}$ of
$0.1489442419$, or both within $10^{-6}$ of its negative. In the positive case
the minimized function is exactly $\ell_C$. In the Horn case, the minimized
function is the probability that the Gaussian rule satisfies the clause, minus
$s_C$. Multiplying this function by $\alpha>0$ and adding $1-\alpha$ gives
$\ell_C$, so both functions have the same minimizers.

\section{Gain from a Gaussian Rounding Window for General CSPs}\label{sec:gain}

\subsection{The Reverse Sweep}\label{sec:updates}

The reverse sweep visits the variables from last to first, so each update can
use the final labels of the neighbors already visited. \Cref{alg:reverse-sweep}
takes the full weighted instance, the score directions $\vect{u}_i$ from a
feasible solution of the SDP in \cref{fig:relaxation}, a fixed rounding
rule, and a window half-width $0<h\le1$. First substitute any labels fixed
by the rule into the predicates, retaining their weights. Number the remaining
variables $1,\ldots,n$ in an order fixed before the Gaussian draw, and write
$t_i$ for their thresholds. The analysis singles out a collection
$\mathcal G$ of binary constraints. For the approximation-ratio guarantee, we
use an SDP solution with interior biases and take $\mathcal G$ to consist of
the binary constraints with slack near zero ($\ell_C<\tau$ for a fixed
$\tau>0$). By \cref{def:near-equality}, this subcollection has regular boundary
geometry on the designated rounding rules.

Write $d_i^{(1)}$ for the total weighted vote of the unary constraints
at $i$. A binary constraint $C=ij$ contributes $w_C\delta_{iC}(y)$ to
the vote at $i$ when $j$ has label $y$.

To compute the forward shift, write
$y_j=\thr_{t_j+s_j}(z_j)$ for the provisional label as a function of the
random score $z_j$. The earlier neighbors contribute the weighted vote
$d_i^-(\vect{y})=\sum_{C=ij,\,j<i}w_C\delta_{iC}(y_j)$ and have total weight
$w_i^-=\sum_{C=ij,\,j<i}w_C$. The forward shift is
\begin{equation}\label{eq:forward-shift}
 s_i=-\frac{h}{w_i^-}\Ex{d_i^-(\vect{y})\mid z_i=t_i}\quad(w_i^->0)\mcom
 \qquad s_i=0\quad(w_i^-=0)\mper
\end{equation}
This expectation fixes $z_i=t_i$ and averages over the orthogonal
Gaussian coordinates using only thresholds already chosen. Since
$|d_i^-(\vect{y})|\le w_i^-$, the shift has magnitude at most $h$. A positive
expected vote lowers the threshold and makes label $+1$ more likely.

\begin{algorithm}[H]
\caption{Reverse sweep with conditional votes}\label{alg:reverse-sweep}
\begin{algorithmic}[1]
  \FOR{$i=1,\ldots,n$}
    \STATE Set $s_i$ by \cref{eq:forward-shift}.
  \ENDFOR
  \STATE Draw a standard Gaussian vector $\vect{g}$.
  \FOR{$i=n,n-1,\ldots,1$}
    \STATE Reveal $z_i=\innerprod{\vect{u}_i}{\vect{g}}$. The revealed information is $\mathcal H_i=\sigma(z_n,\ldots,z_i)$.
    \STATE $y_i\gets \thr_{t_i+s_i}(z_i)$ and $x_i\gets y_i$.
    \IF{$|z_i-(t_i+s_i)|\le2h$}
      \STATE $q_i\gets d_i^{(1)}
        +\sum_{C=ij,\,j>i}w_C\delta_{iC}(x_j)
        +\sum_{C=ij,\,j<i}w_C\Ex{\delta_{iC}(y_j)\mid\mathcal H_i}$.
        \newline{\footnotesize\sffamily\itshape\(\triangleright\)~Use final labels for visited neighbors and average the votes of unvisited neighbors given $\mathcal H_i$.}
      \IF{$q_i\ne0$}
        \STATE $x_i\gets\operatorname{sgn}(q_i)$.
      \ENDIF
    \ENDIF
  \ENDFOR
  \RETURN $\vect{x}$ with the fixed labels restored.
\end{algorithmic}
\end{algorithm}

The sweep uses $q_i$, the total weighted vote at $i$ conditional on
$\mathcal H_i$, to choose the updated label. To evaluate an unvisited
neighbor's contribution, let $K_i$ be the orthogonal projection onto the span
of the revealed directions $\vect{u}_i,\ldots,\vect{u}_n$. The revealed scores
determine $K_i\vect{g}$. The orthogonal component $\vect{g}-K_i\vect{g}$ is
independent of these scores and is standard Gaussian on the orthogonal
complement. For an unvisited neighbor $j<i$, the conditional distribution of
$z_j$ has mean $\mu=\innerprod{\vect{u}_j}{K_i \vect{g}}$ and variance
$\nu^2=\norm{\vect{u}_j-K_i \vect{u}_j}^2$. The conditional probability that
$y_j=-1$ is $\Phi((t_j+s_j-\mu)/\nu)$ if $\nu>0$ and $\1{\mu\le t_j+s_j}$ if
$\nu=0$. An orthonormal basis of the revealed span suffices even when the
directions are linearly dependent.

Write $w_{\mathcal G}=\sum_{C\in\mathcal G}w_C$ for the weight of the
good constraints and $w_{\rm rem}=w_{\rm bin}-w_{\mathcal G}$ for the
remaining binary weight. Substituting fixed labels preserves every payoff
on assignments extending those labels. It leaves $w_{\mathcal G}$ unchanged
because both endpoints of every good constraint have finite thresholds, and
it can only decrease $w_{\rm rem}$. Thus the lower bound below also holds
with the original binary weights. The update at $i$ uses the window of
half-width $2h$ about the shifted threshold $t_i+s_i$. Compare the
sweep output $\vect{x}$ with the unshifted labels
$y_i^0=\thr_{t_i}(z_i)$ on the same draw. For this fixed rule, $a_C$
is the probability that the unshifted assignment satisfies $C$.
On $\mathcal G$, $t_*$ bounds the threshold magnitudes, $r_*$ bounds
the absolute score correlations, and incident oriented normals have
inner product at least $-1+\gamma$.

\begin{samepage}
\begin{theorem}
\label{thm:master}
Fix $t_*,r_*,\gamma$ as in \cref{def:boundary}. There are constants
$c_1,c_2>0$ depending only on these parameters such that, for every
subcollection $\mathcal G$ with this geometry, every $0<h\le1$, every
fixed order, and all deterministic shifts satisfying $|s_i|\le h$,
the reverse pass of \cref{alg:reverse-sweep} with these shifts satisfies
\begin{equation}\label{eq:master}
 \Ex{\val(\vect{x})}-\sum_C w_C a_C
 \ \ge\ (c_1 h/d-c_2 h^2)w_{\mathcal G}-c_2 h w_{\rm rem}\mper
\end{equation}
\end{theorem}
\end{samepage}

At step $i$, the sweep knows all scores $z_i,\ldots,z_n$ and the final labels
of the neighbors already visited. \Cref{lem:crossings} bounds the expected magnitude of
the visible vote. Our main lemma (\cref{lem:payoff}) compares the gain
predicted by this vote with the gain the sweep realizes when neighboring labels
can also change. We keep $\mathcal G$ arbitrary so that baseline slack can
cover the loss on the remaining constraints.

\subsection{A First-Moment Bound for Threshold Votes}

At step $i$, the scores revealed so far generate
$\mathcal H_i=\sigma(z_n,\ldots,z_i)$. The visible vote
$v_i=\Ex{D_i\mid\mathcal H_i}$ is the conditional expectation of the
frozen vote $D_i$. A nonzero $v_i$ favors changing the
baseline label on one half of $[t_i-h,t_i+h]$. Since $v_i$ is independent
of $z_i$, the payoff comparison in \cref{lem:payoff} credits this step
with $\Ex{|v_i|}$ times the smaller Gaussian mass of the two halves,
before charging for interference. At every variable incident to a good
constraint, $t_i$ is bounded, so this mass is $\Omega(h)$. To obtain the
positive $h w_{\mathcal G}/d$ term in \cref{eq:master}, it therefore
suffices to prove $\sum_i\Ex{|v_i|}=\Omega(w_{\mathcal G}/d)$, where the
sum is over those variables.

By \cref{lem:geometry}, $D_i$ is a constant plus weighted halfspace
indicators on $\vect{u}_i^\perp$. At step $i$, the revealed subspace $L$
is the span of the visited score directions projected onto $\vect{u}_i^\perp$.
Conditioning the frozen vote on the Gaussian component in $L$ gives $v_i$.
In the lemma below, we take $J$ to index the good constraints whose other
endpoint has already been visited. Their normals lie in $L$, so each good
constraint contributes to the linear term once, at the second endpoint
visited. In the quadratic error, each
constraint has at most $d$ possible partners per endpoint. Taking the Gaussian
perturbation scale $\lambda$ to be a sufficiently small multiple of $1/d$
therefore gives the required bound after summing over $i$. In the estimate
below, $\tau_*$ bounds the absolute halfspace thresholds and $-1+\eta$ bounds
the inner products of distinct normals from below.

\begin{lemma}[Visible part of the frozen vote]
\label{lem:crossings}
Let $\vect{g}$ be a standard Gaussian vector in a finite-dimensional Euclidean
space and $P$ the orthogonal projection onto a subspace $L$.
Let $\vect{e}_1,\ldots,\vect{e}_m$ be unit normals with
$\innerprod{\vect{e}_j}{\vect{e}_k}\ge-1+\eta$ for $j\ne k$ and $0<\eta\le2$.
For weights $w_j\ge0$, thresholds $|\tau_j|\le\tau_*$, and any constant
$b\in\mathbb R$, let
\[
 D=b+\sum_{j=1}^m w_j
       \1{\innerprod{\vect{e}_j}{\vect{g}}\ge\tau_j}\mper
\]
If $J$ indexes any subset of the normals in $L$, then for every
$0<\lambda\le1/2$,
\begin{equation}\label{eq:crossings}
 2\Ex{\bigl|\Ex{D\mid P\vect{g}}\bigr|}
 \ \ge\ c_{\tau_*}\lambda\sum_{j\in J}w_j
        -k_\eta m\lambda^2\sum_{j=1}^m w_j\mper
\end{equation}
The positive constants $c_{\tau_*}$ and $k_\eta$ depend only on the
indicated parameters.
\end{lemma}

A \emph{crossing} occurs when two samples lie on opposite sides of a
constraint's threshold. Entering its halfspace adds its weight to the vote, and
leaving subtracts it. We perturb the revealed coordinates in opposite
directions at scale $\lambda$ and use the same hidden coordinates for both
samples. The segment between the samples crosses any given revealed boundary
with probability at least $c_{\tau_*}\lambda$. For a given pair of halfspaces,
the probability that the segment enters one and leaves the other is
$O(\lambda^2)$. Thus, whenever $J$ carries positive weight, a small enough
$\lambda$ makes the right-hand side of \cref{eq:crossings} positive.

\begin{proof}
\emph{Couple the two samples.}
Draw independent standard Gaussian vectors $\vect{p},\vect{u}$ on $L$ and an
independent standard Gaussian vector $\vect{h}$ on $L^\perp$. Use $\vect{h}$ as
the hidden component of both samples. Perturb their visible components in
opposite directions by setting
\[
 \vect{m}=\vect{h}+\sqrt{1-\lambda^2}\,\vect{p},
 \qquad \vect{g}^{\pm}=\vect{m}\pm\lambda\vect{u}\mper
\]
In every unit direction in $L$, the two independent summands of the visible
component have variances $1-\lambda^2$ and $\lambda^2$. Each visible component
is therefore standard Gaussian on $L$ and independent of $\vect{h}$, so each
sample has the law of $\vect{g}$.

\emph{Read a sign from the revealed coordinates.}
For $\vect{a}\in L$, write $v_L(\vect{a})$ for the expected vote when
$P\vect{g}=\vect{a}$. We compute it by fixing the component in $L$ at
$\vect{a}$ and averaging over the independent Gaussian component in
$L^\perp$. Thus $v_L(P\vect{g})=\Ex{D\mid P\vect{g}}$.
Follow the segment from $\vect{g}^-$ to $\vect{g}^+$. Write
$\delta_j=1$ if it enters the $j$th halfspace, $-1$ if it leaves, and
$0$ if it stays on the same side. Endpoints lie on a boundary with
probability zero. The change in the total vote is $\sum_jw_j\delta_j$.
Among the crossings in $J$, take the first in a fixed order and let
$s$ be its sign. Set $s=0$ if none crosses. The two projections onto
$L$ determine $s$. Averaging over their common hidden component
therefore gives
\[
 \Ex{s\sum_jw_j\delta_j}
 =\Ex{s\bigl(v_L(P\vect{g}^+)-v_L(P\vect{g}^-)\bigr)}
 \le 2\Ex{|v_L(P\vect{g})|}\mper
\]
Credit each crossing $j\in J$ with $w_j$. A crossing in direction $s$ earns its
weight in the signed sum. A crossing in the opposite direction costs at most
$2w_j$ relative to this credit. It forms an oppositely directed pair with the
crossing that determined $s$. If no index in $J$ crosses, both the credit and
the signed sum vanish. Consequently
\[
 2\Ex{|v_L(P\vect{g})|}
 \ \ge\ \sum_{j\in J}w_j\Pr{\delta_j\ne0}
 -2\sum_{j=1}^m w_j\sum_{k\in J}\Pr{\delta_j\delta_k=-1}\mper
\]
It remains to estimate these two probabilities.

\emph{A single revealed crossing.}
Write $m_j=\innerprod{\vect{e}_j}{\vect{m}}$ and
$u_j=\innerprod{\vect{e}_j}{\vect{u}}$. The endpoint scores are
$m_j\pm\lambda u_j$. They straddle $\tau_j$ precisely when
$|m_j-\tau_j|<\lambda|u_j|$, up to the null boundary event. For $j\in J$,
$u_j$ is standard normal and $m_j$ is an independent normal of variance
$1-\lambda^2$. On the event $1\le|u_j|\le2$, which has fixed positive
probability, the interval for $m_j$ has length at least $2\lambda$ and
lies in $[-\tau_*-1,\tau_*+1]$. The density of $m_j$ is bounded below
there uniformly for $\lambda\le1/2$. Hence
\begin{equation}\label{eq:single-cross}
 \Pr{\delta_j\ne0}\ge c_{\tau_*}\lambda\qquad(j\in J)\mper
\end{equation}

\emph{Two opposite crossings.}
Put $r=\innerprod{\vect{e}_j}{\vect{e}_k}$. If $r=1$, the two normals coincide
and the displacements have the same sign, even if the thresholds differ.
Otherwise $-1<r<1$. Conditional on $\vect{u}$, both crossings require the pair
of midpoint scores $(m_j,m_k)$ to lie in a rectangle of area
$4\lambda^2|u_ju_k|$. The probability is at most this area times the maximum
joint density of the pair.

The midpoint has variance $1-\lambda^2$ on $L$ and variance one on
$L^\perp$. Equivalently, it has the law of a standard Gaussian on the
ambient space scaled by $\sqrt{1-\lambda^2}$ plus independent Gaussian noise on
$L^\perp$ of variance $\lambda^2$. The scaled Gaussian gives the two
scores a joint density bounded by $1/(2\pi(1-\lambda^2)\sqrt{1-r^2})$.
Adding the noise averages translates of this density, preserving the
bound.

The density can be large for nearly parallel normals, but their
displacements are then nearly equal. Opposite crossings require
$u_ju_k<0$, so $4|u_ju_k|\le(u_j-u_k)^2$ on that event, and
\[
 \Ex{(u_j-u_k)^2}
 =\norm{P(\vect{e}_j-\vect{e}_k)}^2
 \le\norm{\vect{e}_j-\vect{e}_k}^2=2(1-r)\mper
\]
Multiplying the density bound by the expected area for opposite
crossings yields
\begin{equation}\label{eq:opposite}
 \Pr{\delta_j\delta_k=-1}
 \le\frac{\lambda^2}{\pi(1-\lambda^2)}
          \sqrt{\frac{1-r}{1+r}}
 \le k'_\eta\lambda^2\mper
\end{equation}
The last bound uses $1+r\ge\eta$ and $\lambda\le1/2$. For $j=k$ or
$r=1$ the crossing directions agree whenever both cross. Substituting
\cref{eq:single-cross,eq:opposite} into the signed sum bound and using
$|J|\le m$ gives \cref{eq:crossings} with $k_\eta=2k'_\eta$.
\end{proof}

\subsection{Comparing the Visible Vote with the Actual Gain}

\Cref{lem:crossings} bounds the expected magnitude of the visible vote
$v_i$. The sweep chooses its label using $q_i$, which includes the votes
cast under the final labels of visited neighbors. To prove \cref{thm:master}, we must show
how much of the gain predicted by $v_i$ the sweep realizes over the
unshifted assignment $\vect{y}^0$.

The inner window is $[t_i-h,t_i+h]$. Let $\omega_i$ be the smaller Gaussian
mass of its two halves, let $\mathcal B_i=\{|z_i-t_i|\le3h\}$ be the event
that the score lies in the enlarged window, and let
$b_i=\sum_{C=ij\notin\mathcal G}w_C$ be the total weight of the remaining
binary constraints at $i$.

A nonzero $v_i$ favors changing the baseline label on one half of the inner
window. The visible vote depends only on the revealed Gaussian component
orthogonal to $\vect{u}_i$, so conditioning on the vote leaves the mass of
that half unchanged. Following its sign therefore gives a predicted gain of
at least $\omega_i\Ex{|v_i|}$. The main lemma shows how
much of that gain survives the actual updates. For a good constraint $C=ij$
with mixed difference $m_C$, the weighted vote at $i$ changes by
$\widehat w_C=w_C|m_C|$ in absolute value when $j$ flips. The constraint incurs
a charge only when \emph{both} endpoints are in their enlarged windows. Their
score correlation is bounded away from $\pm1$, so this event has probability
$O(h^2)$. A remaining binary constraint can incur a charge whenever the current
endpoint is in its enlarged window, an event of probability $O(h)$. At an
endpoint of a good constraint, $\omega_i=\Omega(h)$, so \cref{lem:crossings}
supplies the positive $\Omega(h w_{\mathcal G}/d)$ term in \cref{eq:master}.
For $h$ a sufficiently small multiple of $1/d$, this absorbs the
$O(h^2w_{\mathcal G})$ loss on the good constraints.

\begin{lemma}[Main lemma]\label{lem:payoff}
The expected gain of the reverse sweep satisfies
\begin{equation}\label{eq:payoff}
 \Ex{\val(\vect{x})-\val(\vect{y}^0)}\ \ge\
 \sum_i \omega_i\Ex{|v_i|}
 -4\sum_{C=ij\in\mathcal G}\widehat w_C\Pr{\mathcal B_i\cap \mathcal B_j}
 -2\sum_i b_i\Pr{\mathcal B_i}\mper
\end{equation}
The sum over edges counts each constraint once.
\end{lemma}

\begin{proof}
\emph{Measure the change in sweep order.}
Replace the coordinates of $\vect{y}^0$ by those of $\vect{x}$ in the order
$n,\ldots,1$. The changes $\delta_i$ in satisfied weight telescope to
$\val(\vect{x})-\val(\vect{y}^0)$. If both endpoints of a binary constraint
change, we evaluate the first change with the other endpoint at its baseline
label and the second with the first endpoint at its final label. Let $q_i^0$ be
the expected incident vote in this comparison, conditional on $\mathcal H_i$.
It uses $x_j$ for a visited neighbor and $y_j^0$ for an unvisited one. A flip
from $-1$ to $+1$ earns the incident vote, and the reverse flip earns its
negative. Since $x_i$ and $y_i^0$ are determined by the revealed scores,
\[
 \Ex{\delta_i\mid\mathcal H_i}=\frac{x_i-y_i^0}{2}\,q_i^0\mper
\]
The sweep also uses final labels on visited neighbors. At an unvisited
neighbor $j$, it averages the constraint's vote using the provisional
label $y_j$ in $q_i$, while the payoff comparison uses $y_j^0$ in $q_i^0$.

\emph{Locate the error in the frozen vote.}
Write $\mathcal I_i=\{|z_i-t_i|\le h\}$ for the event that $z_i$ lies in the
inner window. Since $|s_i|\le h$, the update window $\{|z_i-t_i-s_i|\le2h\}$
contains $\mathcal I_i$ and lies inside $\mathcal B_i$. Outside the update
window, the score lies on the same side of both thresholds, so $x_i=y_i^0$ and
$\delta_i=0$.

Inside the update window, freezing $z_i$ moves $z_j$ by
$|r_{ij}(z_i-t_i)|\le3h$. If $|z_j-t_j|>3h$, its frozen label agrees
with the baseline, and $z_j$ is more than $2h$ from its shifted
threshold. Its provisional and final labels thus also equal the
baseline. A good constraint $C=ij$ therefore casts its frozen vote
except possibly on $\mathcal B_j$, where changing $j$'s label alters
that vote by at most $\widehat w_C=w_C|m_C|$. Unary votes are exact.
Each binary constraint omitted from $D_i$ casts a vote of magnitude
at most its weight. Since the update event is known given $\mathcal H_i$,
conditional averaging gives, throughout that window,
\[
 \max\{|q_i-v_i|,|q_i^0-v_i|\}\le\varepsilon_i,\qquad
 \varepsilon_i=b_i+\sum_{C=ij\in\mathcal G}\widehat w_C
                       \Pr{\mathcal B_j\mid\mathcal H_i}\mper
\]

\emph{Compare the chosen label with the frozen prediction.}
For a trial label $x\in\{\pm1\}$, the frozen gain is $(x-y_i^0)v_i/2$, whose
maximum is $(|v_i|-y_i^0v_i)/2$. Inside the update window, the sweep chooses
$x_i$ so that $x_i q_i=|q_i|$. This identity also holds in a tie. If $x_i$
opposes a nonzero $v_i$, the loss relative to the best frozen choice is
$|v_i|$. Then $x_i v_i<0\le x_i q_i$, so $|v_i|\le|q_i-v_i|\le\varepsilon_i$.
Once the label is chosen, evaluating its gain with $q_i^0$ can cost at most
another $\varepsilon_i$, since $|x_i-y_i^0|/2\le1$. We can now keep the
nonnegative predicted gain on the inner window and charge both errors on the
enlarged window. This gives
\[
 \Ex{\delta_i\mid\mathcal H_i}
 \ \ge\ \1{\mathcal I_i}\frac{|v_i|-y_i^0v_i}{2}
          -2\1{\mathcal B_i}\varepsilon_i\mper
\]

\emph{Average over the score.}
Let $P_i$ be the orthogonal projection onto the span of
$\vect{u}_j-r_{ij}\vect{u}_i$ for $j>i$. For each such $j$,
\[
 z_j-r_{ij}z_i
   =\innerprod{\vect{u}_j-r_{ij}\vect{u}_i}{\vect{g}_i^\perp}\mper
\]
Knowing these inner products is equivalent to knowing $P_i\vect{g}_i^\perp$,
so $\mathcal H_i=\sigma(z_i,P_i\vect{g}_i^\perp)$. Both $D_i$ and
$P_i\vect{g}_i^\perp$ depend only on $\vect{g}_i^\perp$, which is independent
of $z_i$. Thus $v_i=\Ex{D_i\mid P_i\vect{g}_i^\perp}$ is independent of $z_i$.
For $v_i>0$, the positive term equals $|v_i|$ on the lower half of the inner
window, and for $v_i<0$ it equals $|v_i|$ on the upper half. Conditioning on
$v_i$ leaves $z_i$ standard Gaussian, so the expected contribution is at least
$\omega_i\Ex{|v_i|}$. For the error term, $\mathcal B_i$ is known at step $i$,
giving
\[
 \Ex{\1{\mathcal B_i}\Pr{\mathcal B_j\mid\mathcal H_i}}
       =\Pr{\mathcal B_i\cap\mathcal B_j}\mper
\]
Each good constraint is charged $2\widehat w_C$ at each endpoint on
the same joint event, giving the coefficient four. The remaining charge
at $i$ is $2b_i\Pr{\mathcal B_i}$. Summing the step gains gives
\cref{eq:payoff}.
\end{proof}

\subsection{Summing the Gains and Using Baseline Slack}

\begin{proof}[Proof of \cref{thm:master}]
\Cref{lem:payoff} credits vertex $i$ with a predicted gain of
$\omega_i\Ex{|v_i|}$ and then subtracts the cost of interference. Early in the
sweep, few neighboring scores have been revealed, so an individual $v_i$ can be
small. We use \cref{lem:crossings} to bound the sum of the expected magnitudes
over the entire sweep.

\emph{Count each good constraint at the second endpoint visited.}
Fix a good constraint $C=ij$ with $i<j$. When the sweep reaches $i$,
it has already revealed $z_j$. The normal $\vect{e}_{iC}$ is a scalar
multiple of $\vect{u}_j-r_{ij}\vect{u}_i$ and hence belongs to the span
of the visited directions projected onto $\vect{u}_i^\perp$. At each
vertex $i$ incident to a good constraint, apply \cref{lem:crossings} on
$\vect{u}_i^\perp$ with this span as the revealed subspace $L$. Take $J$
to consist of the good constraints at $i$ whose
other endpoint has been visited. The conditional vote in the lemma is
then $v_i$, with the unary votes included in its constant term. By
\cref{lem:geometry,def:boundary}, the halfspaces have weights
$\widehat w_C$, thresholds bounded by $\tau_*=2t_*/\sqrt{1-r_*^2}$,
and angle parameter $\eta=\gamma$.

Each good constraint contributes its weight to the single-crossing term once,
when it belongs to $J$. The error term counts its weight at both endpoints
because opposite crossings can cancel at either one. If $m_i$ denotes the
number of good constraints at $i$, a crossing there has at most $m_i\le d$
possible partners. Writing
$\widehat w_{\mathcal G}=\sum_{C\in\mathcal G}\widehat w_C$, we obtain from
\cref{eq:crossings}
\[
 2\sum_{i:m_i>0}\Ex{|v_i|}
 \ge \bigl(c_{\tau_*}\lambda-2k_\gamma d\lambda^2\bigr)
       \widehat w_{\mathcal G}\mper
\]
The factor $d$ in the quadratic error sets the perturbation scale.
Taking $\lambda=k/d$ with $k=\min\{1/2,c_{\tau_*}/(4k_\gamma)\}$
keeps the error at most half the linear term. Hence
$\sum_{i:m_i>0}\Ex{|v_i|}\ge c_{\tau_*}k\widehat w_{\mathcal G}/(4d)$.

\emph{Use the visible vote in the rounding window.}
A nonzero $v_i$ favors changing the baseline label on one half of
$[t_i-h,t_i+h]$. Since $v_i$ is independent of $z_i$, conditioning on
the vote leaves the Gaussian mass of either half unchanged. At a vertex
with $m_i>0$, we have $|t_i|\le t_*$ and $h\le1$, so the density on
both halves is at least $\phi(t_*+1)$. Thus $\omega_i\ge h\phi(t_*+1)$.
Vertices with $m_i=0$ contribute nonnegatively, giving
\[
 \sum_i\omega_i\Ex{|v_i|}
 \ge h\phi(t_*+1)\sum_{i:m_i>0}\Ex{|v_i|}
 \ge \frac{c_{\tau_*}k\phi(t_*+1)}{4}\,
       \frac{h\widehat w_{\mathcal G}}{d}\mper
\]

\emph{Bound the cost of interference.}
On a good constraint $C=ij$, \cref{lem:payoff} charges a loss only if
\emph{both} endpoint scores lie in their enlarged windows. The corresponding
rectangle in the $(z_i,z_j)$-plane has area $(6h)^2$. Since
$|r_{ij}|\le r_*<1$, the joint
Gaussian density is at most $1/(2\pi\sqrt{1-r_*^2})$. The total charge
on good constraints is therefore $O(h^2)\widehat w_{\mathcal G}$.
For a remaining binary constraint, a charge at $i$ requires only
$z_i\in[t_i-3h,t_i+3h]$, an event of probability at most $6h\phi(0)$.
Summing over both endpoints gives a total charge of $O(h)w_{\rm rem}$,
since $\sum_i b_i=2w_{\rm rem}$.

Finally, $w_{\mathcal G}\le\widehat w_{\mathcal G}\le2w_{\mathcal G}$
because $1\le|m_C|\le2$ on a good constraint. Substituting the gain and
the two charges into \cref{eq:payoff}, and using
$\Ex{\val(\vect{y}^0)}=\sum_C w_Ca_C$, gives \cref{eq:master}.
\end{proof}

\subsection{A Criterion for Boolean Constraint Languages}\label{sec:criterion}

We now use the binary constraints with small slack to improve a baseline
approximation ratio. The baseline chooses its rounding rule before the Gaussian
draw and may select a deterministic assignment. For each constraint $C$, let
$\bar a_C$ be its expected payoff over the choice of rule and the rounding. We
apply the reverse sweep on a designated set of Gaussian rules of total
probability at least $\theta>0$. These rules have finite thresholds, and the
algorithm can recognize them when sampled.

\begin{cspdefinition}[Regularity near equality]\label[definition]{def:near-equality}
Fix a finite Boolean constraint language $\mathcal L$, a baseline as above, and
a target ratio $0<\alpha<1$. The language is \emph{regular near equality} for
this baseline at $\alpha$ if there are uniform constants $\tau,\theta>0$ and
$t_*,r_*,\gamma$ as in \cref{def:boundary} with the following property. For
every instance over $\mathcal L$ and every feasible solution of
\cref{fig:relaxation} with $|\xi_i|<1$ for all $i$, both conditions below hold.
\begin{enumerate}[label=\textup{(\roman*)},leftmargin=*]
\item $\ell_C=\bar a_C-\alpha s_C\ge0$ for every constraint $C$.
\item The binary constraints with $\ell_C<\tau$ jointly have regular
boundary geometry with parameters $t_*,r_*,\gamma$ on each designated
Gaussian rule (up to probability zero).
\end{enumerate}
\end{cspdefinition}

The constants are uniform over instances, degrees, and feasible SDP solutions.
At a fixed target $\alpha$, we consider the languages that admit an efficient
baseline with this property. For each of the three disjunction languages, we
take $\alpha$ to be its sharp worst-case ratio. Only pairs of incident binary
constraints enter the angle condition. Unary and constant constraints may have
zero slack.

\paragraph{Checking the equality cases.}
For rules whose thresholds depend only on the endpoint bias and whose mean
payoffs are functions of local configurations, there is a finite-dimensional
way to check the geometric part of the definition. Suppose the designated list
of Gaussian rules is finite. Assume that for some $\tau_0>0$ all binary
configurations of slack below $\tau_0$ lie in a compact set with interior
biases on which the slack and the designated thresholds are continuous. It
suffices to check that $|r_{ij}|<1$ on every tight binary configuration and
that, on every jointly feasible pair of tight incident constraints, the two
oriented normals have inner product strictly greater than $-1$. Indeed, all
thresholds on the compact set are bounded. If there were no uniform correlation
gap near equality, we could choose a sequence of configurations whose slacks
tend to zero and whose absolute score correlations tend to one. A convergent
subsequence would then give a tight configuration with $|r_{ij}|=1$. With that
gap established, the inner product of the normals is continuous in the Gram
entries of $\vect{v}_0,\vect{v}_i,\vect{v}_j,\vect{v}_k$. A sequence of
incident pairs with both slacks tending to zero and inner products tending to
$-1$ would have a convergent subsequence of these Gram matrices, with the same
predicates and rule. Its limit would be a feasible tight pair with opposite
normals. This also covers parallel constraints by allowing $j=k$.

In the next lemma, we select the binary constraints whose expected payoff
exceeds their target by less than $\tau$. The geometry bounds
$(t_*,r_*,\gamma)$ in \cref{def:boundary} must hold for this collection on
every designated rule. Together, these rules have probability at least
$\theta$.

\begin{lemma}[Gain from binary constraints]
\label{lem:binary-weight}
For each binary constraint $C$, fix a target $0\le b_C\le\bar a_C$. Suppose
that for some $\tau>0$ the binary constraints with $\bar a_C-b_C<\tau$
jointly have regular boundary geometry with the same parameters $t_*,r_*,\gamma$
on a designated set of Gaussian rules of total probability at least $\theta>0$. Apply
the reverse sweep on those rules and use the baseline on all other draws.
There is a choice of window half-width $h$ proportional to $1/d$ for which
the resulting assignment satisfies
\[
 \Ex{\val(\vect{x})}\ge
 \sum_{C\ {\rm binary}}w_C b_C+
 \sum_{C\ {\rm unary\ or\ constant}}w_C\bar a_C+
 \frac{\eta w_{\rm bin}}{d^2}\mper
\]
The proportionality constant and the gain coefficient $\eta>0$ depend only on
$t_*,r_*,\gamma,\tau,\theta$.
\end{lemma}

\begin{proof}
Let $\mathcal G$ be the binary constraints with $\bar a_C-b_C<\tau$,
and let $p\ge\theta$ be the probability of the designated rules. We apply
\cref{eq:master} conditional on each designated rule and average over
the choice of rule. Let $b_{\rm tot}$ be the sum of the two target
contributions in the statement. The expected baseline value is at least
$b_{\rm tot}+\tau w_{\rm rem}$. Hence
\[
 \Ex{\val(\vect{x})}\ge b_{\rm tot}+
 p(c_1h/d-c_2h^2)w_{\mathcal G}
       +(\tau-pc_2h)w_{\rm rem}\mper
\]
Take $h=\kappa/d$ with $\kappa=\min\{1,c_1/(2c_2),\tau/(2c_2)\}$. The coefficient of
$w_{\mathcal G}$ is then at least $\theta c_1\kappa/(2d^2)$, and that of
$w_{\rm rem}$ is at least $\tau/2$. Since
$w_{\mathcal G}+w_{\rm rem}=w_{\rm bin}$, the claim follows with
$\eta=\min\{\theta c_1\kappa/2,\tau/2\}$.
\end{proof}

For a supplied SDP solution, $s=\sum_C w_Cs_C$ is its value and
$\xi_i$ is the bias at variable $i$. The constant $\eta$ below is the
gain coefficient from \cref{lem:binary-weight}.

\begin{corollary}
\label{cor:transfer}
Suppose the baseline assigns total probability at least $\theta>0$ to a
designated set of Gaussian rules. If the entire binary collection has regular
boundary geometry with the same parameters $t_*,r_*,\gamma$ on every
designated rule, applying the reverse sweep on those rules and using the
baseline otherwise gives a gain of $\Omega(w_{\rm bin}/d^2)$ over the mean
baseline. The implicit constant depends only on $\theta,t_*,r_*,\gamma$.

For a language regular near equality at $\alpha$, fix a feasible SDP solution
of value $s$ with interior biases ($|\xi_i|<1$ for every variable $i$). Then
the updated assignment satisfies
$\Ex{\val(\vect{x})}\ge\alpha s+\eta w_{\rm bin}/d^2$. If, in addition,
$s\ge\opt-\delta$ for some $\delta\ge0$, the better of $\vect{x}$ and an
assignment maximizing the unary and constant terms has expected value at least
$(\alpha+c/d^2)\opt-\alpha\delta$. Here $\eta$ depends only on the regularity
parameters in \cref{def:near-equality}, and $c$ may also depend on $\alpha$.
\end{corollary}

\begin{proof}
For the gain over the baseline, take $b_C=\bar a_C$ and $\tau=1$ in
\cref{lem:binary-weight}. A single Gaussian rule of probability one
gives \cref{thm:general-rounding}. For the bounds at $\alpha$, the interior
biases let us apply \cref{def:near-equality} to the supplied solution.
Part (i) gives $\bar a_C\ge\alpha s_C$ for every constraint, and
part (ii) supplies the geometry required by \cref{lem:binary-weight}
with $b_C=\alpha s_C$. An assignment maximizing the unary
and constant terms has value at least $\opt-w_{\rm bin}$, since removing
the binary constraints reduces the value of an optimal assignment by at
most $w_{\rm bin}$. If
$w_{\rm bin}\le(1-\alpha)\opt/2$, this already gives $(1+\alpha)\opt/2$.
Otherwise the term $\eta w_{\rm bin}/d^2$ in the first assignment's guarantee
exceeds $\eta(1-\alpha)\opt/(2d^2)$. Both cases give the claim with
$c=(1-\alpha)\min\{1,\eta\}/2$.

For an efficient baseline, discard identically zero predicates and renormalize.
If no constraints remain, every assignment is optimal. Otherwise a uniformly
random assignment shows that $\opt\ge1/4$. Choose $\delta=c/(8d^2)$ so that
$\alpha\delta\le c\opt/(2d^2)$. Solve the SDP to additive accuracy $\delta/2$
and replace its Gram matrix $K$ by $K'=(1-\delta/2)K+(\delta/2)I$. Since $I$ is
feasible, so is $K'$, and its biases $(1-\delta/2)\xi_i$ all lie in $(-1,1)$.
The normalized objective drops by at most $\delta/2$, so the value of $K'$ is
at least $\opt-\delta$. Applying the ratio bound proves
\cref{cor:language-ratio}.
\end{proof}

Suppose a feasible solution satisfies $\bar a_C\ge\alpha s_C$ for every $C$. If
all constraints with $\ell_C<\tau$ are binary and jointly reasonable with
bounded thresholds on the designated rules, \cref{lem:geometry} gives
\cref{def:near-equality}(ii). Every unary and constant constraint then has
slack at least $\tau$. Taking $b_C=\alpha s_C$ in \cref{lem:binary-weight}
gives expected value at least $\alpha s+\min\{\eta,\tau\}/d^2$. The restricted
languages require the more general bound that allows tight unit clauses.

\section{An Inverse-Linear Gain from Aligned Thresholds}
\label{sec:linear-gain}

The reverse sweep allows thresholds of either sign, including zero. For
\cref{thm:aligned-sweep}, the oriented thresholds have a common sign at
each variable and their magnitudes are uniformly bounded away from zero
(\cref{def:aligned-thresholds}). With positive thresholds, every crossing
along an outward ray enters a halfspace. Summing their weights gives a
lower bound on the expected absolute frozen vote that permits a window of
half-width $\Theta(1/\sqrt d)$.

\subsection{Crossings at Aligned Thresholds}

As in \cref{lem:crossings}, we count crossings between two points. Here we
stretch a Gaussian point radially in the span of the incident normals to obtain
the degree bound in \cref{thm:aligned-sweep}. The standard normal density
$\phi$ enters through the probability of crossing each threshold.

\begin{lemma}[Halfspaces with thresholds of one sign]
\label{lem:sat-quadratic-vote}
Let $\vect{g}$ be a standard Gaussian vector and let
$(\vect{e}_j)_{j=1}^m$, $m\ge1$, be unit normals whose span has dimension $k$.
For weights $w_j\ge0$, thresholds $\tau_j$, and a constant $b\in\mathbb R$,
let $D=b+\sum_j w_j\1{\innerprod{\vect{e}_j}{\vect{g}}\ge\tau_j}$.
If all thresholds lie in $[0,\tau_*]$ or all lie in $[-\tau_*,0]$ for some
$\tau_*\ge0$, then
\begin{equation}
\label{eq:aligned-radial}
 \Ex{|D|}\ge\frac{c_*}{\sqrt{k}}\sum_j w_j|\tau_j|\phi(\tau_j)\mcom
\end{equation}
where $c_*>0$ depends only on $\tau_*$.
\end{lemma}

\begin{proof}
Without loss of generality, assume $\tau_j\ge0$ for every $j$. If all
thresholds are nonpositive, $-D$ is almost surely the vote with the same
weights, normals $-\vect{e}_j$, thresholds $-\tau_j$, and constant term
$-b-\sum_j w_j$. This follows by complementing each halfspace, whose
boundary has Gaussian measure zero. Since $|-D|=|D|$ and $\phi$ is even,
the bound for this vote gives the claim for $D$.

\emph{Count the crossings along a ray.}
Write $D(\vect{z})$ for the vote at $\vect{z}$. To eliminate the arbitrary
constant $b$, compare the votes at $\vect{g}$ and at $s\vect{g}$ for $s>1$.
If a score is already at least $\tau\ge0$, stretching it by $s$ keeps the
point in that halfspace. Thus $D(s\vect{g})-D(\vect{g})$ is the total weight
of the crossings near $\vect{g}$ along the segment to $s\vect{g}$. A halfspace
at threshold $\tau$ is crossed when its normal coordinate lies in
$[\tau/s,\tau)$, an interval of Gaussian mass at least
$(1-1/s)\tau\phi(\tau)$.

The crossing interval grows with $s$. What limits the stretch is that we
must also control $|D(s\vect{g})|$ using the original Gaussian law.
Projecting onto the span of the normals preserves every score. Identify
this span with $\mathbb R^k$ and write $\vect{g}$ for the projected standard
Gaussian. At $\vect{y}=s\vect{x}$, the log density of $s\vect{g}$ relative
to that of $\vect{g}$ is
\[
 k\left(\frac{s^2-1}{2}-\log s\right)
 +\frac{s^2-1}{2}\bigl(\|\vect{x}\|^2-k\bigr)\mper
\]
For $1<s\le2$ the first term is $O(k(s-1)^2)$, and the squared Gaussian
radius has standard deviation $\sqrt{2k}$ about its mean $k$. We therefore
take $s=1+1/\sqrt{k}$. At this scale the log density ratio is bounded above
whenever $\|\vect{x}\|^2\le k+O(\sqrt{k})$.

To keep a fixed fraction of each crossing, allow $\tau_*^2$ for the squared
normal coordinate. Conditional on that coordinate, the squared length in its
orthogonal complement still has mean $k-1$ and variance $2(k-1)$. Hence,
by Chebyshev's inequality, at least half the mass of each crossing lies in
$A=\{\|\vect{g}\|^2\le k+3\sqrt{k}+\tau_*^2\}$. For $\vect{x}$ in this ball
the density ratio at $s\vect{x}$ is at most a constant $C_{\tau_*}$ depending
only on $\tau_*$. Integrating
the density over points $\vect{y}\in B$ whose preimages $\vect{y}/s$ lie in
the ball gives, for every Borel set $B\subseteq\mathbb R^k$,
\[
 \Pr{A\cap\{s\vect{g}\in B\}}\le C_{\tau_*}\Pr{\vect{g}\in B}\mper
\]
Summing the crossing weights on $A$ and using this comparison at the
stretched endpoint gives
\[
 \tfrac12(1-1/s)\sum_j w_j\tau_j\phi(\tau_j)
 \le \Ex{\1{A}\bigl(D(s\vect{g})-D(\vect{g})\bigr)}
 \le (1+C_{\tau_*})\Ex{|D|}\mper
\]
Since $1-1/s=1/(\sqrt{k}+1)\ge1/(2\sqrt{k})$, this proves
\cref{eq:aligned-radial} with $c_*=1/(4(1+C_{\tau_*}))$.
\end{proof}

\subsection{Simultaneous Updates and Baseline Slack}
\label{sec:aligned-updates}

For a selected constraint $C$ incident to $i$, $\tau_{iC}$ is its
oriented threshold after freezing the score at $i$. The bounds
$\tau_0\le|\tau_{iC}|\le\tau_*$ from \cref{def:aligned-thresholds}
make the vote estimate in \cref{lem:sat-quadratic-vote} uniform over the
selected constraints. On the omitted binary constraints, the parameter
$\tau$ in the following theorem measures the baseline payoff margin.
Independent activation preserves a fixed fraction of that margin.

\alignedThresholdTheorem*

\begin{proof}
Fix a designated Gaussian rule. By \cref{lem:sat-quadratic-vote}, a
variable of total selected halfspace weight $w_i$ has expected absolute
frozen vote $\Omega(w_i/\sqrt d)$. In a window of half-width $0<h\le1$,
an active variable earns the magnitude of a nonzero vote on the half
where its baseline label opposes the vote. At a selected endpoint,
either half has Gaussian mass $\Omega(h)$. With activation probability $p$,
the predicted gain per unit of selected weight is therefore
$\Omega(ph/\sqrt d)$. Interference requires both scores to lie in their
windows and at least one endpoint to be active, so its expected cost on
the same scale is $O(ph^2)$. Taking $h$ to be a sufficiently small
multiple of $1/\sqrt d$ leaves a gain of $\Omega(p/d)$ per unit weight.
With this vote estimate in place, the remaining steps for the selected
constraints follow the proof of \cref{thm:master}. For completeness, we
give the details below, including how independent activation preserves
slack on the remaining constraints.

\emph{Estimate the frozen votes.}
Write $\vect{y}^0$ for the baseline assignment on this rule. At a variable
whose label is deterministic, keep that label fixed and put $D_i=0$. At every
other variable $i$, compute the full frozen vote $D_i$ from this draw, summing
the original unary votes at $i$ and the votes of incident binary constraints in
$\mathcal G$.
Every selected $C=ij$ has two finite thresholds and $|r_{ij}|<1$,
so the halfspace calculation in the proof of \cref{lem:geometry} applies.
Its indicator has weight $\widehat w_C=w_C|m_C|$. Let
$w_i=\sum_{C=ij\in\mathcal G}\widehat w_C$. The unary votes contribute
only to the constant term of $D_i$.

Fix $i$ incident to a selected constraint. It has at most $d$ selected
neighbors $j$, so their score vectors $\vect{u}_j$ together with
$\vect{u}_i$ span at most $d+1$ dimensions. Projecting $\vect{g}$ onto
this span preserves $z_i$ and each
$z_j$, hence also the local baseline labels and the frozen scores
$z_j^{[i]}=z_j-r_{ij}(z_i-t_i)$. Each oriented normal $\vect{e}_{iC}$ is
a multiple of $\vect{u}_j-r_{ij}\vect{u}_i$, so it lies in the part of
this span orthogonal to $\vect{u}_i$. The normals that determine $D_i$
therefore span at most $d$ dimensions.

By \cref{def:aligned-thresholds}, the selected thresholds at $i$ have one
sign and satisfy $\tau_0\le|\tau_{iC}|\le\tau_*$. Applying
\cref{eq:aligned-radial} with $k\le d$ gives
\[
 \Ex{|D_i|}
 \ge\frac{c_*}{\sqrt d}\sum_{C=ij\in\mathcal G}
          \widehat w_C|\tau_{iC}|\phi(\tau_{iC})
 \ge c_*\tau_0\phi(\tau_*)\,\frac{w_i}{\sqrt d}\mper
\]
Here $c_*$ depends only on $\tau_*$. If $i$ has no selected constraint,
$w_i=0$ and the last bound holds as well.

\emph{Use the frozen vote in the rounding window.}
Activate each variable independently with probability $p$, using randomness
independent of the baseline. On a designated rule, set
$x_i=\operatorname{sgn}(D_i)$ at each active vertex with a finite threshold
whose score satisfies $|z_i-t_i|\le h$ and whose vote $D_i$ is nonzero.
Otherwise keep $x_i=y_i^0$. Write $\Delta_i=(x_i-y_i^0)/2$, so the
predicted gain at $i$ is $\Delta_iD_i$. If $D_i\ne0$, the predicted gain is
$|D_i|$ when $i$ is active and its score lies in the half-window where the
baseline label opposes the vote. The vote $D_i$ depends only on
$\vect{g}_i^\perp$, which is independent of $z_i$. At an endpoint of a selected
constraint, $|t_i|\le t_*$, so each half has Gaussian mass at least
$h\phi(t_*+1)$ even after conditioning on $D_i$. At every other vertex, the
predicted gain is nonnegative. Writing
$w_{\mathcal G}=\sum_{C\in\mathcal G}w_C$ and summing over the vertices gives
\[
 \sum_i\Ex{\Delta_iD_i}\ge c_1p\,\frac{h}{\sqrt d}\,w_{\mathcal G}\mcom
\]
since $\sum_i w_i=2\sum_{C\in\mathcal G}\widehat w_C\ge2w_{\mathcal G}$.

\emph{Bound the cost of interference.}
On a selected constraint $C=ij$, the actual change in payoff is
\[
 f_C(x_i,x_j)-f_C(y_i^0,y_j^0)
 =\Delta_i\delta_{iC}(y_j^0)+\Delta_j\delta_{jC}(y_i^0)
   +m_C\Delta_i\Delta_j\mper
\]
To predict the first term, we use the label at $j$ obtained by freezing $z_i$
at $t_i$. When $i$ flips, this changes $z_j$ by at most $|r_{ij}|h$. The frozen
label at $j$ can therefore differ from $y_j^0$ only if
$|z_j-t_j|\le |r_{ij}|h$. Thus an error requires activation at $i$ and both
scores in their windows. The same reasoning applies at $j$. The last term can
be nonzero only if both endpoints are active and both scores lie in their
windows. Since $|r_{ij}|\le r_*<1$, the joint density of $(z_i,z_j)$ is
uniformly bounded, and the two windows define a square of side $2h$ with
probability $O(h^2)$. As $|m_C|\le2$, the two freezing errors cost $O(ph^2)$
per unit weight and the interaction costs $O(p^2h^2)$. Unary gains are exact.
Consequently, the expected gain on the selected and unary constraints,
conditional on this rule, is at least
\[
 p\left(c_1\frac{h}{\sqrt d}-c_2 h^2\right)w_{\mathcal G}\mper
\]

\emph{Choose the wider window.}
Per unit of selected weight, \cref{eq:master} balances $h/d$ against
$h^2$, giving a window of half-width $\Theta(1/d)$. Here, after
factoring out $p$, we balance $h/\sqrt d$ against the same $h^2$
cost. The window can therefore be wider by a factor of $\sqrt d$. Take
$h=\kappa/\sqrt d$ with
$0<\kappa\le\min\{1,c_1/(2c_2)\}$. Then
$c_2h^2\le c_1h/(2\sqrt d)$, leaving a gain of
$\Omega(p w_{\mathcal G}/d)$ on each designated rule. We return the
baseline assignment on every other rule. Averaging gives a gain of
$\Omega(\theta p w_{\mathcal G}/d)$ over the mean baseline
contribution of the selected and unary constraints.

\emph{Retain the baseline slack on the remaining constraints.}
Choose $p=\tau/(4(1+\tau))$. For an omitted binary constraint $C$,
both endpoints are inactive with probability $(1-p)^2$, independently
of the baseline rule and assignment. This event preserves its payoff,
so $\Ex{f_C(\vect{x}_C)}\ge(1-p)^2\bar a_C$. Under the relative slack
hypothesis, $\bar a_C\ge b_C+\tau s_C$ and $b_C\le s_C$ give
\[
 (1-p)^2\bar a_C-b_C
 \ge\bigl((1-p)^2\tau-(2p-p^2)\bigr)s_C
 \ge \tfrac{\tau}{2}s_C\mper
\]
Under absolute slack, the expected decrease is at most
$(2p-p^2)\bar a_C\le2p$, leaving at least $\tau/2$ above $b_C$.

The expected baseline payoff meets every target, and constant constraints keep
their payoff. Since $s_C\le1$, the selected weight $w_{\mathcal G}$ is at least
the total weighted SDP payoff of the selected constraints. Under relative
slack, each omitted binary constraint retains a weighted margin of at least
$\tau w_Cs_C/2$. Since $d\ge1$, the two bounds give the claimed inverse-linear
gain on the total binary SDP payoff. Under absolute slack the retained margin
is at least $\tau w_C/2$, giving a gain of $cw_{\rm bin}/d$ above the total
target.
\end{proof}

\paragraph{From the SDP payoff to an approximation ratio.}
Fix $0<\alpha<1$ and take $b_C=\alpha s_C$ in
\cref{thm:aligned-sweep}. Write $s=\sum_Cw_Cs_C$ and
$s_{\rm bin}=\sum_{C\ {\rm binary}}w_Cs_C$. The relative-margin bound
gives expected value at least $\alpha s+cs_{\rm bin}/d$. The
absolute-margin bound gives the same inequality since
$w_{\rm bin}\ge s_{\rm bin}$. An assignment maximizing the unary and
constant constraints has value at least $s-s_{\rm bin}$. At each
variable the SDP averages over its two labels, and the maximizing
labels can be chosen independently. Take the better of this assignment
and the updated one. If $s_{\rm bin}\ge(1-\alpha)s/2$, the gain is at least
$c(1-\alpha)s/(2d)$. Otherwise the unary and constant objective alone
is at least $(1+\alpha)s/2$. In either case the expected value is at
least $(\alpha+c'/d)s$ with $c'=(1-\alpha)\min\{1,c\}/2$.

For a guarantee that holds throughout a language, the hypotheses must hold with
uniform constants for every feasible SDP solution with interior biases. We must
also be able to implement the baseline, recognize the designated rules, and
identify the selected constraints in polynomial time. Discard identically zero
predicates and normalize the remaining weights as in \cref{cor:transfer}.
Unless the instance is empty, $\opt\ge1/4$. Solving the SDP to additive
accuracy $c_0/d$ and mixing with the identity by a sufficiently small amount of
order $1/d$ gives interior biases and retains a ratio of $\alpha+\Omega(1/d)$
for sufficiently small $c_0>0$.

\paragraph{Scope of the condition.}
Reasonableness gives the correlation bound in \cref{def:aligned-thresholds}(i).
The lower bound $\tau_0>0$ in part~(ii) is a separate requirement. For
example, place equality constraints on the edges of a star. Choose the
center sign uniformly. Given that sign, choose the leaf signs independently,
matching the center with probability $3/4$. The moments are feasible for the
SDP. Every bias is zero, every edge has product moment and score
correlation $1/2$, and every mixed difference is positive, so the
collection is reasonable. If all rounding thresholds are zero, the
oriented thresholds are also zero. Stretching a Gaussian point along its
ray then preserves every halfspace indicator, and the right-hand side of
\cref{eq:aligned-radial} vanishes. The inverse-square guarantee of
\cref{thm:general-rounding} still applies.

\section{Disjunction Languages}
\label{sec:sat}

We apply \cref{thm:aligned-sweep} to Max 2-SAT and to the two restricted clause
languages. In each language, the binary clauses near equality lie in fixed
neighborhoods where the oriented thresholds align. The update also uses unit
clauses, even when their slack is zero.

\subsection{The Thresholds of a Signed Clause}

To check alignment for a signed clause $C=ij$, express the thresholds and the
score correlation in literal coordinates. Put $t_i^{\rm lit}=\epsilon_i t_i$,
$t_j^{\rm lit}=\epsilon_j t_j$, and $r_{\rm lit}=\epsilon_i\epsilon_j r_{ij}$, where
$\epsilon_i,\epsilon_j$ are the literal signs. The mixed difference in the
original variable coordinates is $m_C=-\epsilon_i\epsilon_j$.
\Cref{def:aligned-thresholds} gives
\[
 \tau_{iC}=-\epsilon_i
  \frac{t_j^{\rm lit}-r_{\rm lit}t_i^{\rm lit}}{\sqrt{1-r_{\rm lit}^2}}\mper
\]
Suppose the two literal thresholds have the same sign, their magnitudes are
bounded above and below by positive constants, and $r_{\rm lit}<0$ is bounded
away from $-1$. Then $t_j^{\rm lit}-r_{\rm lit}t_i^{\rm lit}$ has the sign of
$t_i^{\rm lit}$ and magnitude at least $|t_j^{\rm lit}|$. The oriented
threshold at $i$ therefore has the sign of
$-t_i$ for either polarity of the literal, and its magnitude is bounded above
and away from zero. The same holds at $j$.

\subsection{Max 2-SAT}

\begin{proof}[Proof of \cref{thm:main}]
We verify alignment for the LLZ rule of \cref{sec:clause-baselines}. At the two
tight triples $(a,b,\rho)=(\pm u,\pm u,-1+2u)$, the signs agree and $u>0$. The
LLZ threshold map is odd and nonzero at these biases. In literal coordinates,
the centered correlation is $-(1-u)/(1+u)\in(-1,0)$. Small fixed rational boxes
about the tight triples therefore have the required threshold and correlation
bounds. For feasible solutions with interior biases, the LLZ separation bound
in \cref{sec:clause-baselines} gives a uniform positive lower bound on the
slack of every binary clause outside the two boxes. Select precisely the
clauses in the boxes. The preceding calculation gives aligned thresholds, and
the baseline satisfies $\bar a_C\ge\alpha_{\rm LLZ}s_C$ on all constraints,
including units and tautologies. The absolute-slack bound in
\cref{thm:aligned-sweep} now applies. Converting the gain to an approximation
ratio as in \cref{sec:aligned-updates} proves the theorem.
\end{proof}

\subsection{Horn Clauses and Positive Binary Disjunctions}
\label{sec:restricted}

\begin{proof}[Proof of \cref{thm:restricted}]
Fix either language and let $\alpha$ be its ratio in
\cref{sec:clause-baselines}. The baseline rounds a variable of bias $b$ to
false with probability $\alpha(1+b)/2$. Hence negative unit clauses have zero
slack, and positive unit clauses and tautologies have slack $1-\alpha$.

We first show that a binary clause has uniformly positive slack whenever either
literal is nearly deterministic in the SDP solution. Let $u,v$ be the SDP
probabilities that the literals of a binary clause are true. In $\mathcal L_+$
the rounded truth probabilities are $1-\alpha+\alpha u$ and
$1-\alpha+\alpha v$. In $\mathcal L_{\rm H}$ the Gaussian rule preserves both
literal marginals, and the all-true assignment satisfies the clause. In either
case, $\bar a_C\ge1-\alpha+\alpha\max\{u,v\}$ and $s_C\le\min\{1,u+v\}$. It
follows that
\[
 \ell_C\ge 1-\alpha-\alpha\min\{u,1-u\}\mper
\]
To see this, use the lower bound on $\bar a_C$ with $v$ and the upper bound
$s_C\le u+v$ to obtain $\ell_C\ge 1-\alpha-\alpha u$. Using $u$ and $s_C\le1$
instead gives $\ell_C\ge 1-\alpha-\alpha(1-u)$. The same argument applies with
$v$ in place of $u$. Thus the slack is at least $(1-\alpha)/2$ near either
deterministic endpoint.

On the remaining compact set of local configurations, the baseline payoff is
continuous, including at score correlation $\pm1$. To see continuity, write the
scores as $Z_1$ and $rZ_1+\sqrt{1-r^2}Z_2$, where $Z_1,Z_2$ are independent
standard normal variables. Along any convergent sequence of parameters, each
indicator converges almost surely because the limiting score equals its
threshold with probability zero. By \cref{sec:clause-baselines}, the two
literal biases at every tight triple have the same sign, are bounded away from
zero, and have magnitude less than $0.2$. The product moment lies between
$-0.72$ and $-0.62$, so the centered correlation is negative. Correlation $-1$
would force the two nonconstant labels to be opposites almost surely, giving
product moment $-1$. Thus the centered correlations of the tight triples lie in
a compact subinterval of $(-1,0)$.

For the Horn language, designate the Gaussian branch of the baseline. This
branch has probability $\alpha$ and uses the odd, bias-preserving map. The two
literal thresholds at a tight triple have the same nonzero sign. For positive
disjunctions, designate the unique Gaussian rule. The tight biases are both
close to $-0.1824168$, and the rounded bias $-1+\alpha(1+b)$ is negative there.
Again the literal thresholds have the same nonzero sign. In both languages,
these properties persist throughout sufficiently small fixed rational
neighborhoods of the tight triples. Select the binary clauses in these
neighborhoods. Compactness and the endpoint bound give a uniform positive lower
bound on the slack of the omitted binary clauses. For the selected clauses, the
signed-clause calculation gives aligned thresholds on the designated rule. We
can therefore apply the absolute-slack bound in \cref{thm:aligned-sweep}, even
in the presence of tight negative unit clauses. The conversion in
\cref{sec:aligned-updates} gives the claimed ratios.
\end{proof}

\phantomsection
\section*{Acknowledgements}
\addcontentsline{toc}{section}{Acknowledgements}
The author thanks Johan H{\aa}stad for suggesting the bounded-degree
Max 2-SAT problem in 2022 and recommending Mikael Flor\'{e}n's thesis on
Max-Cut~\cite{FlorenThesis}, and Aaron Potechin for pointing out the paper
on the tight approximability of Max 2-SAT~\cite{BHZ23}.

\phantomsection
\section*{AI Disclosure}
\addcontentsline{toc}{section}{AI Disclosure}
OpenAI's ChatGPT helped analyze the author's proposed threshold shifts and
reverse sweep, including the comparison of the frozen vote with the actual
gain and the bound on opposite crossings. It also helped derive the radial
estimate for aligned thresholds and generated the initial draft of the
manuscript (including this disclosure) under the author's direction.

\printbibliography[heading=bibintoc]
\end{document}